\documentclass[twocolumn, superscriptaddress, showpacs,
 amsmath,amssymb,
 aps,
 prab
]{revtex4-2}

\usepackage{graphicx}
\usepackage{dcolumn}
\usepackage{bm}

\newcommand{\m}{\text{m}}
\newcommand{\mev}{\text{MeV}}
\newcommand{\gev}{\text{GeV}}

\newcommand{\gevm}{\gev/\m}

\newcommand{\cm}{\text{cm}}
\newcommand{\cc}{\text{cm}^{-3}}
\newcommand{\mum}{\text{$\mu$}\m}

\newcommand{\joule}{\text{J}}
\newcommand{\fs}{\text{fs}}
\newcommand{\mm}{\text{mm}}

\newcommand{\pc}{\text{pC}}
\newcommand{\rad}{\text{rad}}
\newcommand{\mrad}{\text{mrad}}
\newcommand{\wpercm}{\text{W}/\text{cm}^{2}}
\newcommand{\tw}{\text{TW}}
\newcommand{\fwhm}{\text{FWHM}}

\begin{document}

\preprint{APS/123-QED}

\title{A compact, all-optical positron production and collection scheme}

\author{Davide Terzani}
\email{dterzani@lbl.gov}
\affiliation{%
 Lawrence Berkeley National Laboratory, Berkeley, CA 94720, USA
}
\author{Carlo Benedetti}
\affiliation{%
 Lawrence Berkeley National Laboratory, Berkeley, CA 94720, USA
}
\author{Stepan S. Bulanov}
\affiliation{%
 Lawrence Berkeley National Laboratory, Berkeley, CA 94720, USA
}
\author{Carl B. Schroeder}
\affiliation{%
 Lawrence Berkeley National Laboratory, Berkeley, CA 94720, USA
}
\affiliation{Department of Nuclear Engineering, University of California, Berkeley, CA 94720, USA}
\author{Eric Esarey}
\affiliation{%
 Lawrence Berkeley National Laboratory, Berkeley, CA 94720, USA
}%

\date{\today}

\begin{abstract}

In this paper we discuss a compact, laser-plasma-based scheme for the generation of positron beams suitable to be implemented in an all-optical setup.
A laser-plasma-accelerated electron beam hits a solid target producing electron-positron pairs via bremsstrahlung.
The back of the target serves as a plasma mirror to in-couple a laser pulse into a plasma stage located right after the mirror where the laser drives a plasma wave (or wakefield).
By properly choosing the delay between the laser and the electron beam the positrons produced in the target can be trapped in the wakefield,
where they are focused and accelerated during the transport,
resulting in a collimated beam.
This approach minimizes the ballistic propagation time and enhances the trapping efficiency.
The system can be used as an injector of positron beams and
has potential applications in the development of a future, compact, plasma-based electron-positron linear collider.

\end{abstract}

\maketitle


\section{\label{sec:introduction}Introduction}

Over the past few decades, a great interest has grown around Laser-Plasma Accelerators (LPAs)
owing to the possibility of generating accelerating gradients that are several orders
of magnitude larger than that obtainable in conventional,
radio-frequency-based accelerators~\cite{esarey_physics_2009,hooker_developments_2013}.
This makes them attractive candidates as compact drivers for
free-electron-lasers~\cite{wang_free-electron_2021,labat_seeded_2023}
or for a high-energy
linear collider~\cite{benedetti_whitepaper_2022, benedetti_linear_2022}.
In an LPA, an intense laser pulse propagates in a plasma
and separates the electrons from the background ions via the action of the ponderomotive force generating a plasma wave (or wakefield).
A particle beam injected into the wakefield at a suitable phase
can be focused and accelerated to ultra-high energies.
Laser-plasma acceleration of electrons has been demonstrated~\cite{faure_laserplasma_2004, geddes_high-quality_2004, mangles_monoenergetic_2004, wang_quasi-monoenergetic_2013, gonsalves_petawatt_2019}
and the research focus is now shifting towards applications,
high-quality beam production, and shot-to-shot reproducibility~\cite{maier_decoding_2020}.
On the other hand, high-quality, high-efficiency plasma-based positron acceleration
remains a critical challenge that needs to be addressed
in order to enable the design of a plasma-based $e^+e^-$ collider~\cite{musumeci_positron_2022}.

The development of a plasma-based scheme for positron acceleration presents several difficulties, and the development of schemes allowing for high-gradient and high-efficiency acceleration represents an area of active research.
Recently, many plasma-based positron acceleration schemes have been proposed~\cite{hue_efficiency_2021, silva_stable_2021, zhou_high_2021, diederichs_high-quality_2020},
but the production of a positron beam with collider relevant parameters
(i.e., high-energy and ultra-low emittance)
has still not been demonstrated.
In addition, there are currently no facilities that can produce positron beams
for experiments, which limits the research possibilities.
In order to start future plasma-based positron acceleration experiments,
researchers are currently investigating several techniques to generate a positron beam,
including pair production caused by either the passage
or the generation of an electron beam in a solid target~
\cite{chen_relativistic_2009, chen_relativistic_2010, sarri_laser-driven_2013, sarri_generation_2015, fujii_positron_2019, streeter_narrow_2022},
and pair creation from the interaction of an electron beam with an ultra-intense laser pulse in the strong-field
quantum electrodynamics regime~\cite{lobet_generation_2017, vranic_multi-gev_2018, gonoskov_charged_2022, zhu_dense_2016, zhao_all-optical_2022, martinez_creation_2023}.
However, trapping a positron beam of significant charge remains challenging.
Accumulation rings generate very long particle beams $\left(\sim \mm\text{-scale}\right)$ that are not suitable to be injected into a plasma wave since the characteristic wake size, which depends on the plasma density, is on the order of $10-100\,\mum$
for plasma densities in the range $10^{19}-10^{17}\,\cc$.
The goal of a positron beam injector is to be able to produce a high-charge and short positron beam in a single shot.
It is desirable to use laser-plasma accelerated electron
beams as a source
to generate positrons as they are available in compact setups
and they naturally comply with the requirement to generate an ultra-short and high-current particle beam.

\begin{figure}[th]
    \centering
    \includegraphics[width=8.6cm]{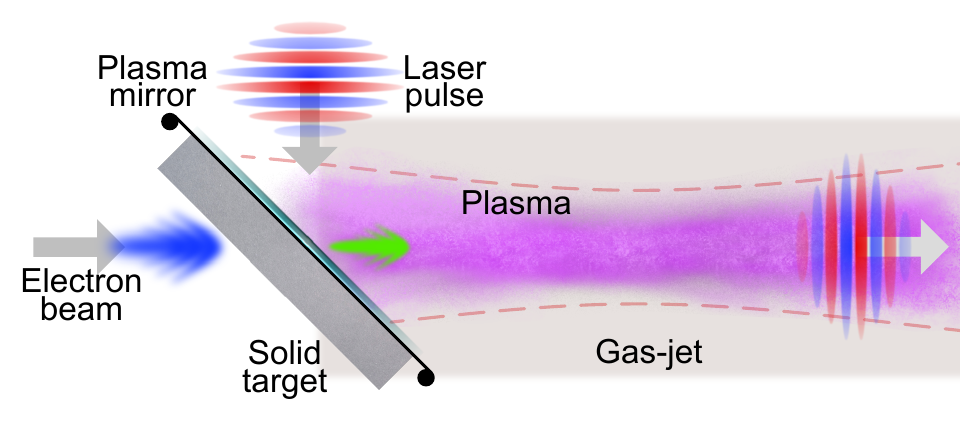}
    \caption{Design of the positron generation and collection scheme.
    An electron beam (coming from the left) hits a thick tungsten target and $e^+e^-$
    pairs are generated via bremsstrahlung.
    A laser pulse (coming from the top) impinges on the back surface of the target, covered, for instance, with a VHS tape, and is reflected by the plasma mirror to the right, where it excites a plasma wave.
    Positrons exiting the target (green) are trapped and accelerated in the wake.
    The primary beam and secondary electrons after the target are not represented in the picture for the sake of clarity.}
    \label{fig:Expsetup}
\end{figure}

In this paper we propose a compact generation and collection scheme for positrons entirely based on readily available laser and plasma technology
that maximizes the positron beam charge trapped in a plasma wave.
This scheme is based on the high-energy electron beam
interaction with a solid density foil to produce positrons,
which are subsequently captured and accelerated in the plasma wave generated by a laser pulse that is in-coupled in an
underdense plasma using a plasma mirror.
In Fig.~\ref{fig:Expsetup} we show a schematic of the proposed scheme.
A $10\,\gev$ LPA-generated electron beam (also called primary beam)
impinges upon a thick tungsten target rotated by $45^\circ$
around the vertical axis
and placed close to a supersonic gas-jet.
During the interaction with the solid target,
the electron beam produces $e^+e^-$ pairs via bremsstrahlung
that exit the target along with the primary beam.
The particle cloud produced in such a process is characterized
by quasi-charge neutrality
(i.e., $Q_{e-}\simeq Q_{e+},$ where $Q_{e-}$ and $Q_{e+}$ are the charge
of the produced electrons and positrons, respectively)
and a large RMS divergence.
For a target size of several radiation lengths,
defined as the distance traveled by a particle when its
energy is reduced by a factor $1/e\simeq0.37$~\cite{chao_handbook_2013},
the primary beam is heavily perturbed during the interaction,
resulting in a substantial energy loss and divergence increase when it reaches the back of the target.
A laser pulse
is reflected by a plasma mirror~\cite{sokollik_tapedrive_2010, shaw_reflectance_2016} located on the back side of the target
along the propagation line of the incoming particle beam,
ionizes the gas ejected from a gas-jet,
and excites a (linear) plasma wave in the plasma.
By properly delaying the beam arrival, the positrons extracted from
the target are directly injected into an accelerating and focusing plasma wave phase. This phase of the plasma wave is defocusing for electrons and, hence, all the primary and secondary electrons are defocused and lost.
Any transport distance between the positron source and the plasma wave
contributes to the loss of a significant fraction of the positron charge produced \cite{amorim_design_2023}.
In fact, the large divergence and broad energy spectrum of particles generated
via bremsstrahlung makes it challenging for these particles
to be transported using conventional focusing optics.
For this reason, it is of paramount importance to take into account a realistic placement
of the experimental components, as failing to do so
could overestimate the trapping efficiency of the scheme~\cite{sahai_quasimonoenergetic_2018}.
Conversely, the possibility, unlocked by the plasma mirror,
to generate an accelerating and focusing plasma wakefield right after the target
maximizes the amount of trapped charge.

This article is organized as follows.
In Sec.~\ref{sec:G4Simulation},
we describe the bremsstrahlung process and the numerical tool used to model it.
In Sec.~\ref{sec:Lineartrapping},
we discuss the parameters of the LPA stage.
Simulation results for various laser and plasma parameters are discussed in Sec.~\ref{sec:results},
with a particular emphasis on how one can maximize the positron charge in the final beam.
In Sec.~\ref{sec:conclusions}, we draw conclusions for this work.

\section{\label{sec:G4Simulation} Montecarlo simulation of the beam-target interaction}

We modeled the interaction of an LPA generated electron beam with a thick tungsten target
using a custom code based on the Geant4 toolkit
\cite{agostinelli_geant4simulation_2003, allison_geant4_2006, allison_recent_2016}.
The electron beam parameters are the ones expected in an LPA stage driven by the BELLA PW laser~\cite{benedetti_infrno_2018}.
We consider a $E_b=10\,\gev$ electron beam with $0.5\%$ energy spread, normalized emittance $\varepsilon_n=1\,\mum$, and RMS divergence
$\sigma_\theta = 0.2\,\mrad$.
We take into account the beam divergence in the Geant4 simulations
as it has an impact on the final positron beam quality.
\begin{figure}[ht]
    \centering
    \includegraphics[width=8.6cm]{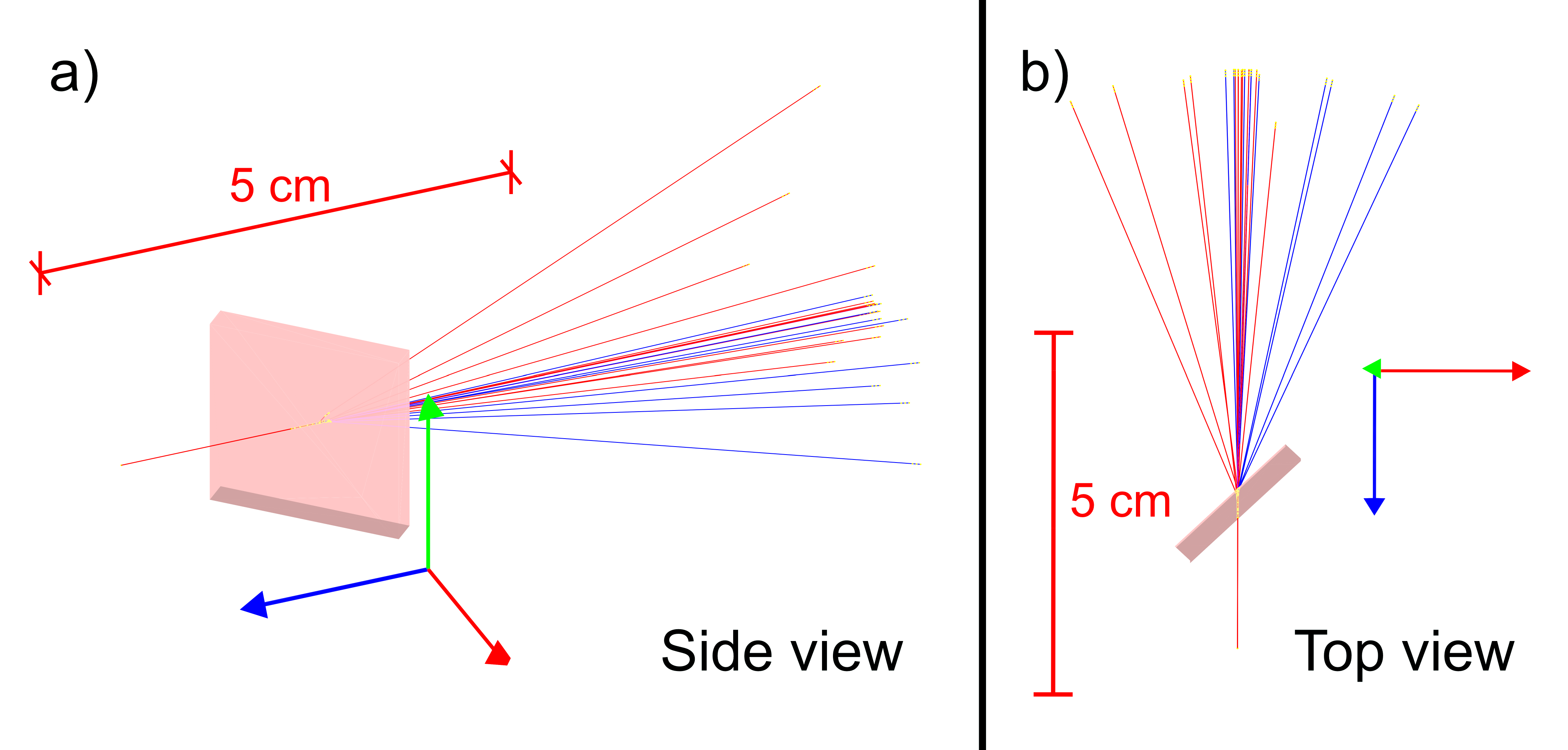}
    \caption{Beam-target collision setup simulated in Geant4.
    The incoming electron beam travels along the $-z$ axis (blue arrow)
    and hits one side of the tungsten target, rotated by $45^\circ$.
    Secondary particles are produced and exit the target from the other side.
    Electrons trajectories are shown in red, positrons trajectories in blue.
    a) Side view. b) Top view.}
    \label{fig:G4setup}
\end{figure}
The beam is modeled via a ballistic propagation of the particles that travel
along the $-z$ direction for $2\,\cm$ from the end of the LPA stage to the tungsten target
(radiation length $X_0=3.5\,\mm$),
that is rotated by $45^\circ$ around the $y$ axis, as is shown in Fig.~\ref{fig:G4setup}.

An high-energy electron can generate positrons in a two-step process when traversing
an high-Z solid target.
First, the electron loses energy by emitting a high-energy photon
in the direction of propagation when scattering on a heavy nucleus.
Then, when such photon recoils on one of the nuclei of the target it may decay into a lepton pair,
in particular a $e^+e^-$ pair with the highest probability.
Such mechanism is called Bethe-Heitler process and it is the most probable channel for
the production of positrons using an high-energy electron beam interacting with a thick high-Z target.
An analogous process, named Trident process, predicts the emission of a $e^+e^-$ pair
when an electron scatters on the nucleus.
However, it was shown \cite{gryaznykh_estimates_1998, nakashima_numerical_2002}
that for ultrarelativistic electron beam energies and large target thicknesses, i.e.,
$\Delta \psi \gtrsim X_0$, the probability of such process is negligible compared to the Bethe-Heitler.

We performed several simulations varying the thickness of the target in the range
$2.8\,\mm\leq \Delta \psi \leq 7.7\,\mm$, i.e., $0.8 X_0\leq \Delta \psi \leq 2.2 X_0$
in terms of the radiation length.
We point out that the thickness $\Delta \psi$
is the thickness of the tungsten target,
and the effective distance traveled by one particle inside the
target is $\sim\sqrt{2}\Delta\psi$.
The 6D phase space of the particles exiting the back of the target
was then used as an input to simulate their dynamics in the wakefield
generated by the laser pulse reflected by the plasma mirror.
We emphasize that the positron generation process in the solid converter is linear
with respect to the incoming electron beam charge
and the number of initial particles used in Geant4
is only relevant for a statistical purpose.
Therefore, in this paper, rather than referring to a positron charge,
we will use a dimensionless number, namely the capturing efficiency $Y$ defined as the ratio of the charge of the positrons, $Q_{e+}$,
over the charge of the incoming electron beam, $Q_0$.
Such number defines the capturing efficiency of the scheme.
The definition here presented is the combination (product)
of the efficiencies of the two sub-processes involved.
In fact, particles are generated in the target with an efficiency $Y_0$
that depends on both the beam and target parameters.
Of all the positrons exiting the target, only a fraction $Y_1$ have
the suitable initial conditions that allow trapping into the wakefield.
Thus, we define the capturing efficiency as $Y=Y_0Y_1$.
We point out that the overall performance of the scheme is not necessarily optimal when both the number of generated positrons and the wakefield amplitude at the capturing point are independently optimized.
Due to the bremsstrahlung process, an increase in the overall number of
positrons corresponds to a degradation of the quality of their distribution.
Therefore, the fraction of the produced positron that is trapped into the plasma is not independent from the characteristics of the target itself.
A definition of capturing efficiency
that takes into account both the charge conversion into the solid target and the trapping power of the wakefield, enables the optimization of the overall process, that would otherwise be more challenging if the two separate stages were to be addressed.

The cloud of secondary particles that is produced in the target is
characterized by the same temporal structure as that of the incoming electron beam.
The transverse size of the secondary particle distribution increases with the target thickness due to multiple scattering and,
for the values of $\Delta \psi$ considered for our setup,
is in the range $10\,\mum \lesssim \sigma_r \lesssim 100 \,\mum,$
while the RMS divergence is usually on the order of $\sigma_\theta \simeq 0.1\,\rad$.

We point out the importance of modeling the positron production using a Monte Carlo code.
Particle generation via bremsstrahlung in thick targets is an highly complex process and, to date, no analytical models for the final particle distributions are available.
Simulating realistic initial beam distribution and positron production exiting the solid target is the only approach that guarantees meaningful results.
For instance, the assumption considered in \cite{sahai_quasimonoenergetic_2018}
of an initial relativistic anisotropic Maxwellian distribution inflates the final performance of the scheme as it models the positron beam having parameters more advantageous for subsequent capture and acceleration than can be acquired in experiments.

In the next section we will present simulations of the dynamics of the produced positrons into the wakefield
and the optimization process that maximizes the trapped positron charge.

\section{\label{sec:Lineartrapping} Definition of laser and plasma parameters}

Particles generated in an electromagnetic shower typically present an exponential-like energy spectrum and a very large divergence due to the multiple scattering they undergo passing through the solid target. For this reason, particle capturing in conventional systems is done using tapered solenoidal magnetic fields that reduce the final beam divergence by adiabatically increasing the transverse beam size up to several $\mm$ \cite{chao_handbook_2013}.
However, the transverse acceptance of a plasma accelerator stage is limited by the waist of the laser driver,
typically $w_0\lesssim 100\,\mum$.
The capturing efficiency of such technique is therefore drastically reduced
and alternatives must be considered to inject positrons into an LPA stage.

In order to mitigate the particle loss due to the large production angles,
ideally the plasma stage should be located in the immediate vicinity of the positron source,
so that the produced particles are subject to an accelerating and focusing
field right after leaving the solid target.
We therefore consider using a plasma mirror placed on the back of the target (realized, e.g., by means of a tape drive) so that an high intensity laser pulse can be reflected from the target and
generate a plasma in its proximity.
In fact, plasma mirrors reflect substantial amount ($>70\%$)
of the incoming pulse energy with an intensity
$I \simeq 10^{16}\,\wpercm$~\cite{scott_optimization_2015, sokollik_tapedrive_2010, shaw_reflectance_2016},
where we can express $I\,\left[\wpercm\right] \simeq 1.4\times 10^{18}a^2/\lambda^2_0\,[\mum]$
as a function of the peak normalized laser strength $a=eA/m_ec^2$
($A$ is the peak laser vector potential,
$e$ is the unit charge, $m_e$ is the electron mass,
and $c$ is the speed of light in vacuum)
and the laser wavelength $\lambda_0$.
A laser pulse coming from a $90^\circ$ angle, as shown in Fig.~\ref{fig:Expsetup},
impinging on the plasma mirror is reflected along the propagation axis of
the incoming electron beam.
The gas expelled from a gas-jet nozzle, placed after the solid target,
is ionized by the pulse, a plasma is formed,
and a wakefield is generated if the laser strength and pulse duration are such that
$a\sim 1$ and
$T_\fwhm\left[\fs\right]\sim 42/\sqrt{n_0\left[10^{18}\cc\right]}$,
where $n_0$ is the plasma density.
If the delay between the laser pulse and the electron beam is properly tuned,
it is possible to trap the positrons produced in the solid target,
while all the electrons are deflected by the defocusing transverse fields.

Assuming a laser waist at focus $w_0=85\,\mum$,
yielding a Rayleigh range $Z_R=\pi w_0^2/\lambda_0 = 2.8\,\cm$ for $\lambda_0=0.8$ $\mum$,
and a pulse intensity on the plasma mirror $I_{PM} = 3\times 10^{16}\,\wpercm$,
the laser focal point is positioned at about 4.5 Rayleigh ranges (corresponding to $12.5\,\cm$) from the mirror, and the laser pulse intensity at focus is $I_0 = 6.6 \times 10^{17}\,\wpercm$,
which corresponds to normalized laser strength at focus $a_0=0.55$.
We consider a plasma having number density $n_0=1\times 10^{17}\,\cc$ and a double-Gaussian laser
 with Full-Width-at-Half-Maximum (FWHM) duration $T_{\fwhm} \simeq 80\,\fs$.
 Using this set of parameters, the required laser power and energy are, respectively,
 $P = 73\,\tw$ and $\mathcal{E}=6.2\,\joule$.

\subsection{\label{sec:simulationscheme} Modeling the dynamics of particles in a linear wakefield}

A laser pulse with intensity $I_0 < 10^{18}\,\wpercm$ excites a plasma
wave in the linear regime~\cite{gorbunov_excitation_1987, esarey_physics_2009}.
Such regime is favorable for positron capturing because
a quarter of the plasma wave is accelerating and focusing for positrons.
LPAs operating in the nonlinear regime provide, in principle, larger accelerating gradients than that operating in the linear regime. However, in the nonlinear regime the wake region suitable for positron acceleration is small, and, hence, positron trapping becomes challenging.



We point out that the concept introduced in this paper is intended as an injection and collection schemes for positrons. Once the positron beam is formed, more advanced positron acceleration techniques, such as hollow channels \cite{zhou_high_2021} or plasma columns \cite{diederichs_positron_2019}, could be used to further boost the beam energy.

Assuming the space charge effects from the beam to be negligible,
particle dynamics in a linear wakefield can be modeled using a
particle tracker code that evaluates the analytical solution of the
plasma fields.
We built a particle tracker that
computes the evolution of the initial particle phase space
subject to the linear wakefield generated by a linearly-polarized, double-Gaussian laser pulse
assuming a perfect reflection from the plasma mirror and negligible transient fields.
The laser longitudinal profile is
$\Pi \left(\zeta\right) = \exp \left(-\zeta^2/2L^2 \right)$,
where $L = c T_{\fwhm}/2\sqrt{\log(2)}$ is the laser duration,
$\zeta = z-c\beta_g t$ is the comoving coordinate,
$c\beta_g$ is the group velocity of the laser pulse in the plasma,
and $\omega_p = ck_p = \sqrt{4\pi e^2n_0/m_e}$ is the plasma frequency.
Behind the laser driver, the accelerating and focusing fields are, respectively,
\begin{multline}
    \label{eq:acclinearfield}
    E_z(r, \zeta, t) = \frac{a^2\left(t\right)}{2}\Theta\left(L\right)\\\times\exp\left[-2r^2/w^2\left(t\right)\right]
    \cos\left(k_p\zeta + \varphi_0\right),
\end{multline}
\begin{multline}
    F_r(r, \zeta, t) = E_r(r, \zeta, t) - B_{\phi}(r, \zeta, t) = \\
    \label{eq:foclinearfield}
    -2 a^2\left(t\right)\frac{r}{k_pw^2\left(t\right)}\Theta\left(L\right)
    \exp\left[-2r^2/w^2\left(t\right)\right]\sin\left(k_p\zeta + \varphi_0\right),
\end{multline}
where $r^2 = x^2 + y^2$,
$\varphi_0$ the initial wakefield phase and
$\Theta \left(L\right) = \sqrt{\pi}k_pL\exp \left(-k_p^2L^2/4 \right)/2$.
The electric and magnetic fields are normalized to the cold wavebreaking limit $E_0=m_ec^2k_p/e$.
In the linear regime, the group velocity of the laser pulse is
$\beta_g = \sqrt{1 - \omega_p^2/\omega_0^2-2c^2/\omega_0^2w_0^2}$ \cite{esarey_physics_2009},
where $\omega_0$ is the laser frequency.
Assuming that the ratio of the laser power over the critical power
$P/P_c = \left(k_p w_0 a_0\right)^2/32 \ll 1$,
the laser pulse follows the Rayleigh diffraction,
with a normalized strength $a\left(z\right)$ and waist $w\left(z\right)$
given respectively by
\begin{align}
\label{eq:rayleighA}
&a\left(z\right) = a_0/\sqrt{1 + \left(z-z_0\right)^2/Z_R^2},\\
\label{eq:rayleighW}
&w\left(z\right)=w_0\sqrt{1 + \left(z-z_0\right)^2/Z_R^2},
\end{align}
with $z_0$ the location of the laser focus.
The field \mbox{Eqs.(\ref{eq:acclinearfield}-\ref{eq:foclinearfield})} are evaluated on each particle position that is evolved using a 4th-order Runge-Kutta temporal integrator.
Particles are propagated in the plasma for $D = 25.5\, \cm\simeq9Z_R$.

We point out that, since we are interested in studying the positron capturing
in the linear regime, the laser strength is limited to values $a_0 \lesssim 0.6$,
after which the linear wakefield approximation starts to deviate from the nonlinear result.
Study of the performance of this scheme with more intense lasers, where the laser-plasma interaction enters the (mildly) non-linear regime, requires a numerical expression for the wake structure. In such mildly-nonlinear regimes, there might exist a favorable working point
where the positron capturing efficiency is increased compared to the linear regime. We anticipate carrying this exploration and analysis in a future work.

\section{\label{sec:results} Simulation results}

\subsection{\label{sec:Linearregime} Unguided laser pulse}
We computed the evolution of the
phase space of the primary beam and of the secondary positrons and electrons
in the linear wakefield generated by the laser pulse for several values of the target thickness.
For every choice of laser and plasma parameters, the phase $\varphi_0$  was chosen to maximize the final number of captured positrons.
We then introduced a metric that determines an optimal thickness
by counting the number of positrons that at the end of the simulation
 gained at least $\Delta E = 50\,\mev$.
As discussed in Sec.~\ref{sec:G4Simulation}, the efficiency of the scheme is given by $Y=Q_{e+}/Q_0$,
and the actual final positron charge scales linearly with the primary beam charge
as long as beamloading effects are negligible.
\begin{figure}[ht]
    \centering
    \includegraphics[width=8.6cm]{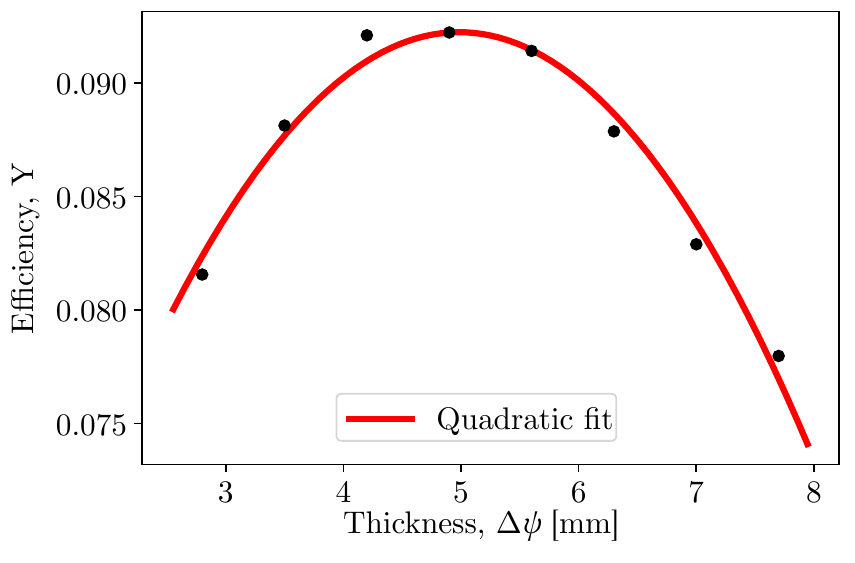}
    \caption{Capturing efficiency of the scheme in function of the thickness of the solid target.
    Positrons are counted at the end of the propagation length $D = 25.5\,\cm$,
    by selecting only particles that gained an energy $\Delta E\geq 50\,\mev$.
    The efficiency $Y=Q_{e+}/Q_0$ is defined as the ratio of the measured positron charge
    over the initial electron beam charge.
    The efficiency reaches its maximum $Y\simeq 9.2\times 10^{-2}$ at $\Delta \psi = 4.9\,\mm$;
    however, the efficiency varies slowly over target thicknesses of several mm.}
    \label{fig:Yield}
\end{figure}

In Fig.~\ref{fig:Yield} we show the efficiency $Y$ versus the target thickness.
There is a maximum around $\Delta \psi = 4.9 \,\mm$,
where the positron charge amounts to $Q_{e+}\simeq 9.2\times 10^{-2} Q_0$,
although the efficiency is weakly dependent on the thickness
in the range $4\,\mm\lesssim \Delta \psi \lesssim 6\,\mm$.

For any given incoming beam parameters, the optimal target thickness is determined by two competing processes.
A thicker target yields more positrons as long as the their absorption probability remains low, i.e., $\Delta \psi \lesssim 5 X_0$ for $E=10\,\gev$, at the same time determining an higher final RMS divergence due to the multiple scattering.
On the other hand, a smaller thickness produces higher quality positron beams with a reduced charge.
An optimal working point can therefore be found in between the two extremes, when an high number of positron is produced with a low enough divergence that allows the particle to be trapped into the wakefield without escaping from the sides.
\begin{figure}[ht]
    \centering
    \includegraphics[width=8.6cm]{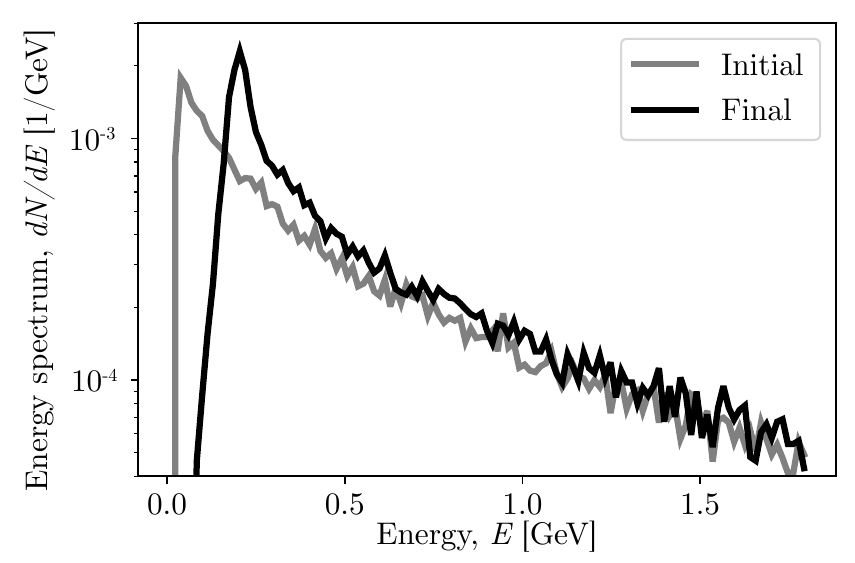}
    \caption{Comparison of the energy spectrum of the positrons right after they leave the tungsten converter (gray) and after the propagation into the plasma (black) having total length $D=25.5\,\cm$.
    We can see a net energy gain that shifts the energy peak from $\sim 40\,\mev$ to $\sim 200\, \mev$.
    To compute the spectra we only consider particles with a minimum total energy gain $\Delta E \geq 50\,\mev$.}
    \label{fig:Initialfinalspectrum}
\end{figure}

In the following, we choose $\Delta \psi=4.9\,\mm$ and analyze the corresponding dynamics of the secondary particles in the wakefield.
Note that all the electrons, either from the primary beam
or produced in the target, are deflected away by the defocusing transverse force of the plasma wakefield,
so none are detected within the plasma wave at the end of the simulation.
On the other hand, a significant fraction of the positrons experience a focusing and accelerating
field, with a net energy gain that is visible in the difference
between the initial and final energy spectra depicted in Fig.~\ref{fig:Initialfinalspectrum}.

It is of interest to look at a single positron energy slice rather
than the whole spectrum, because we can optimize a subsequent magnetic transport line for the accelerated positrons.
To minimize the chromatic effects of the transport line on the beam we
limit the energy spread of the slice to $5\%$.
This is an arbitrary choice aimed at finding a balance between the final beam charge and the requirements of a transport line.
Thus, we selected the $5\%$ energy slice of the final spectrum that contains the maximum positron charge and, for the rest of the paper, we will refer to those particles as a positron beam.
\begin{figure}[ht]
    \centering
    \includegraphics[width=8.6cm]{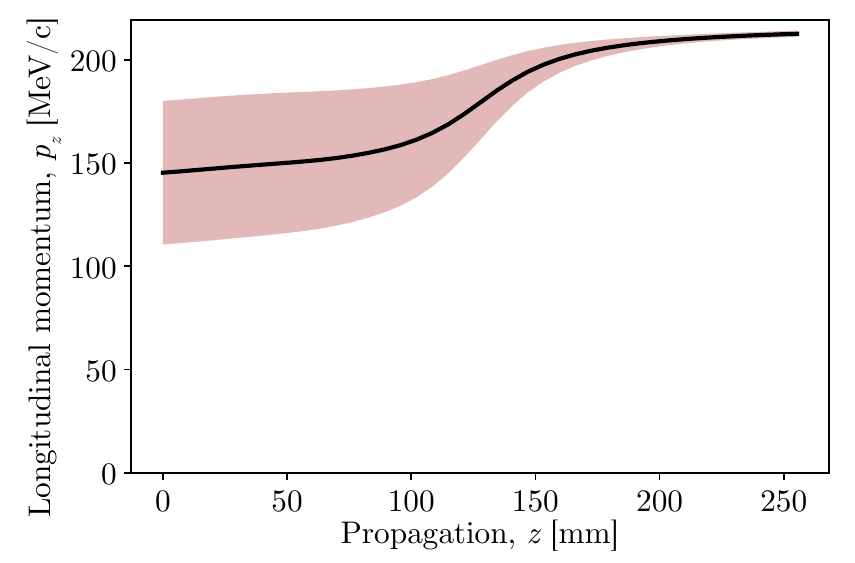}
    \caption{Evolution of the longitudinal momentum of the positrons with a final energy
    $E_0=205\pm5\,\mev$.
    The shaded red region contains the $p_z\pm \sigma_{pz}/2$ values.
    This particular set of particles experiences an average accelerating gradient $E_g > 0.2\,\gevm$.}
    \label{fig:Longmomevolution}
\end{figure}
Considering only positrons with final energies $E_0=205\pm5\,\mev$,
we obtain $Y=8\times 10^{-3}$, e.g., a positron charge $Q_{e+}=0.8\,\pc$ for an initial electron beam charge $Q_0 = 100\,\pc$.
We point out that this value is orders of magnitude higher than the ones achieved by
other LPA-based schemes discussed in the literature~\cite{sarri_laser-driven_2013, martinez_creation_2023, streeter_narrow_2022},
and it is obtained employing currently available PW-class laser technology.
The main reason of such boost in the capturing efficiency of our scheme is the spatial overlapping of the plasma stage with the bremsstrahlung-produced positrons
that is enabled by the use of the plasma mirror.
In Fig.~\ref{fig:Longmomevolution} we show the evolution
of the longitudinal momentum of the positron beam.
Throughout the propagation, the energy gain of the particles is $\Delta E\simeq 50\,\mev$, demonstrating an average accelerating gradients $E_g > 0.2\,\gevm$.

It is possible to boost the accelerating gradient by operating in a regime where the laser
pulse self-focuses during the propagation.
The maximum amplitude of a laser pulse impinging on a plasma mirror is limited by the technology of the mirror itself.
On the other hand, we can freely increase the laser waist and make use of the pulse self-focusing
to enable higher intensities at focus.
For a given distance between the plasma mirror and the focal point, the capturing efficiency scales as
$Y \propto w_{PM}^2$, where $w_{PM}$ is the laser waist at the plasma mirror,
for a laser size smaller that the size of the particle beam,
and it saturates to a maximum value when the laser size is larger than the particle beam size.
In the case under consideration, the laser impinging on the mirror is much larger than the positron cloud
at the back of the target, therefore we do not expect any increase in capturing efficiency
when increasing the laser waist.
However, future plasma mirror technologies could enable reflection of laser pulses with higher intensities,
allowing the focal point to be moved closer to particle source
and improving the capturing efficiency.

In addition to the laser parameters considered in Sec.~\ref{sec:simulationscheme}, as an additional example we also simulated the evolution of a laser having duration $T_{\fwhm}=80\,\fs$,
waist at focus $w_0=120\,\mum$ and strength $a_0=0.6$
in a uniform plasma of density $n_0=1 \times 10^{17}\,\cc$
using the fluid modality of the code INF\&RNO \cite{benedetti_efficient_2010, benedetti_accurate_2017}.
For these parameters, pulse self-focussing is not negligible, although we verified that, throughout the evolution, the generated wakefield is
reproduced by the linear formulas in \mbox{Eqs.(\ref{eq:acclinearfield}) and (\ref{eq:foclinearfield})} within a $< 5\%$ error.
The final energy slice $E_0=438\pm11\,\mev$,
contains a charge $Q_{e+}/Q_0=8\times10^{-3}$.
Compared to the case presented previously,
the higher pulse intensity at focus results in an higher average accelerating gradient that reaches $E_g > 0.6\,\gevm$.
During the beam evolution, the captured beam normalized emittance saturates to a final value
$\varepsilon_n \simeq 90\,\mum$ and is kept constant throughout the propagation.
In fact, the beam spot size is much smaller than the laser waist, thus particles
experience a linear focusing field which preserves emittance.

\subsection{\label{sec:guidedpulse} Increasing the interaction length using a plasma channel}
In the configuration discussed in the previous section the positron energy gain was limited by laser diffraction. The beam energy gain can be increased by guiding the laser over distances much longer that its Rayleigh diffraction length.
Here we consider the performance of the scheme when a parabolic plasma channel starting at the location of the laser focus is used. Within the plasma channel the transverse density profile is given by $n(r)/n_0 = 1 + \alpha r^2$, with the matched value of the channel depth $\alpha=\alpha_M=4/k_p^2w_0^4$. 
Hence, the laser waist evolves according to Eq.~\eqref{eq:rayleighW}
when propagating from the plasma mirror to the focal point,
and it remains constant to the focal value $w_0=85\,\mum$ once in the channel.
The positron beam is trapped into the wakefield undergoing a prolonged acceleration phase, thus reaching higher energies that are only limited by the beam dephasing
(in this regime the laser depletion length is of the order of $\sim 10^2\,\m$ \cite{benedetti_pulse_2015}).
\begin{figure}[ht]
    \centering
    \includegraphics[width=8.6cm]{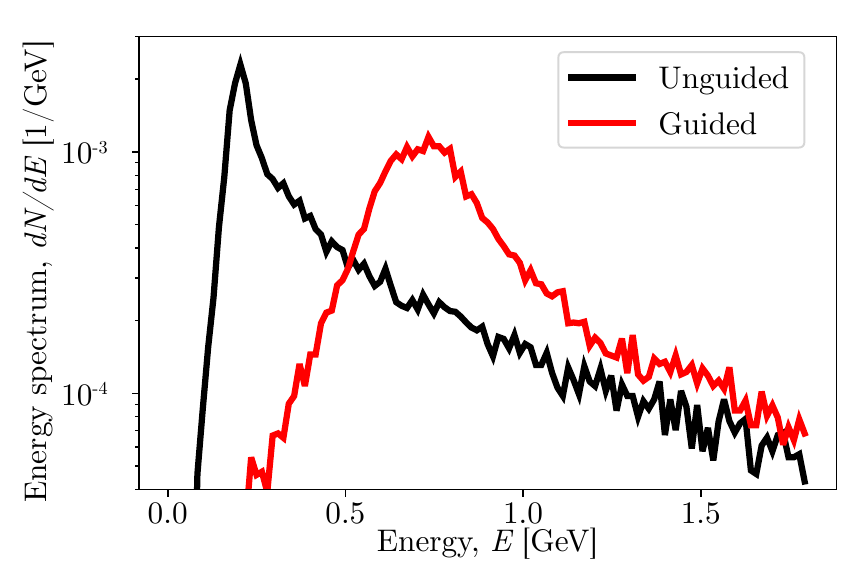}
    \caption{Comparison between the final energy spectra obtained respectively in the
    unguided and in the guided case.
    The final particle energy is considerably increased owing to the longer propagation
    in the plasma $\left(D=51\,\cm\right)$ and the higher accelerating gradient.}
    \label{fig:finalspectracomparison}
\end{figure}
In Fig.~\ref{fig:finalspectracomparison}
we compare the final energy spectrum obtained
propagating the same laser pulse defined in Sec.~\ref{sec:simulationscheme} ($a_0=0.55$, $w_0=85\,\mum$ and $T_\fwhm=80\,\fs$)
 in a uniform plasma (unguided, black line)
and in a parabolic plasma channel (guided, red line), respectively.
It can be seen that in the latter case the energy gain is greatly increased owing to the longer interaction length $D=51\,\cm$.
\begin{figure}[ht]
    \centering
    \includegraphics[width=8.6cm]{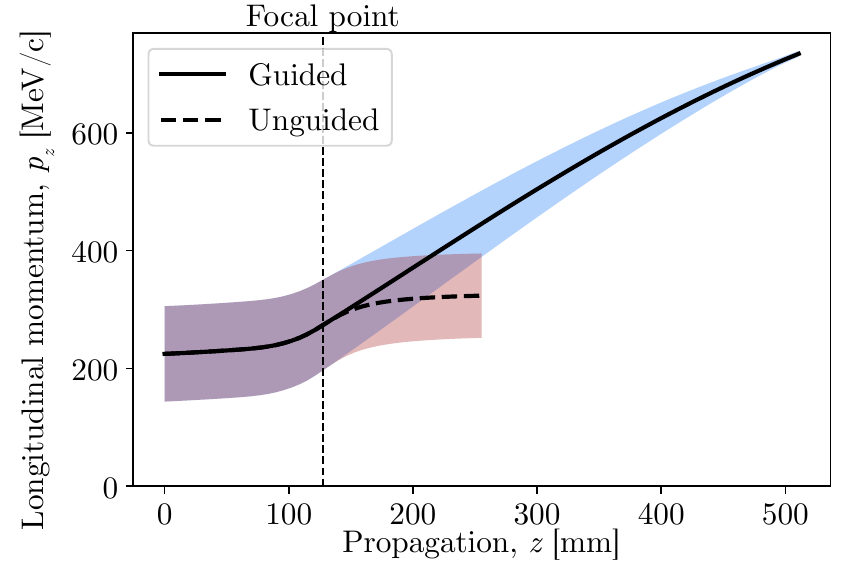}
    \caption{Comparison between the evolution of the longitudinal momentum of the same set of particles in the unguided and the guided case.
    The vertical black dashed line represents the laser focal point,
    where the plasma channel starts.
    Particles are selected such that their final energy in the guided case is $E_0=734 \pm 18\,\mev$.
    The shaded regions contain the $p_z \pm \sigma_{pz}/2$ values.
    In the guided case, the accelerating gradient remains almost constant throughout the evolution
    owing to the laser driver not diffracting.
    Eventually, the maximum achievable energy is limited by the beam dephasing.
    This particular selection of particles contains a charge $Q_{e+}=8\times 10^{-3}Q_0$.}
    \label{fig:Longmomcomparison}
\end{figure}
We selected a $5\%$-energy spread slice around the final energy $E_0=734\,\mev$, measuring an efficiency $Y=8\times 10^{-3}$,
and compared the dynamics of the same set of particles in the guided and
in the unguided case.
As it is shown in Fig.~\ref{fig:Longmomcomparison},
after the laser focal point the accelerating gradient remains close to its maximum value
and the positron beam accelerates until it reaches dephasing,
which for this particular set of parameters happens farther than the simulated $51\,\cm$,
and, as expected, the final energy is higher than in the unguided case.
Overall, we can estimate an average accelerating gradient $E_g\simeq1\,\gevm$.
\begin{figure}[ht]
    \centering
    \includegraphics[width=8.6cm]{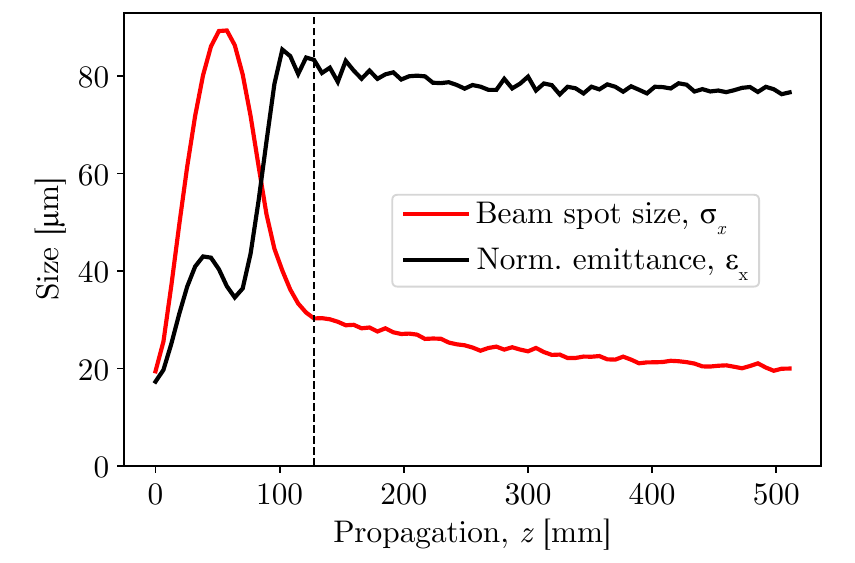}
    \caption{Evolution of the normalized emittance and the spot size of the positron beam when a matched plasma channel is used to guide the laser pulse.
    The vertical black dashed line represents the laser focal point,
    where the plasma channel starts.
    After a transient phase, the emittance saturates to $\varepsilon_n \simeq 80\,\mum$
    and the spot size slowly decreases to $\sigma_x \simeq 20\,\mum$.}
    \label{fig:Emittanceevolutionguided}
\end{figure}
Furthermore, channeling the laser pulse confines the beam for a longer distance lowering the final positron divergence.
In Fig.~\ref{fig:Emittanceevolutionguided},
we show the evolution of the normalized emittance and of the spot size of the positron beam trapped in the wakefield in the guided case.
After a transient phase during which the emittance increases
due to phase mixing and nonlinearities in the focusing fields,
its value $\varepsilon_n \simeq 80\,\mum$ is preserved during the propagation.
Conversely, after the focal point the spot size $\sigma_x$ slowly decreases as the energy increases reaching $\sigma_x\simeq 20\,\mum$.

\section{\label{sec:conclusions} Conclusions}

We presented a positron generation and collection scheme that is based on currently
available LPA technology and that does not rely on any conventional focusing optics.
The setup includes a $45^\circ$ rotated solid target to which a plasma mirror is attached.
This particular configuration maximizes the number of positrons that are trapped in the
wakefield by minimizing the distance between the positron source and the wakefield that traps them.

For a laser pulse exciting a wakefield in the linear regime,
namely characterized by $a_0=0.55$ and $w_0=85\,\mum$, 
travelling in a uniform plasma of density $n_0=1\times 10^{17}\,\cc$,
our modeling shows how the wakefield generated by such pulse
is able to accelerate a number of positrons up to $Q_{e+}=8\times 10^{-3}Q_0$ within a $5\%$ energy spread around the final energy $E_0=205\,\mev$.
Over a propagation length of $D=25.5\,\cm$,
the beam experiences an average accelerating gradient $E_g \approx 0.2\,\gevm$.
By increasing the laser waist and strength to
$w_0=120\,\mum$ and $a_0=0.6$, respectively, the average accelerating
gradient reaches $E_g \approx 0.6\,\gevm$.
Finally, we showed that a plasma channel placed at the laser focal point improves the performance by
increasing the total interaction length and
confining the positron beam into the maximum accelerating field throughout the propagation.
We have shown how the average accelerating gradient using a plasma channel reaches $E_g \approx 1\,\gevm$,
and yields a final positron energy on the order of $E_0 \sim 1\,\gev$,
with a final value limited by the beam dephasing.

The use of a plasma mirror significantly boosts the capturing efficiency compared to analogous positron capturing systems.
Additional improvement to the final performance of the scheme is possible. 
For example, improved plasma mirror technology could enable reflection of laser pulses even closer to the focal point,
increasing the capturing efficiency of the wakefield at the particle production source.
The capabilities of this scheme could also be explored in mildly nonlinear regimes,
where the stronger accelerating and focusing power of the wakefield could yield a higher capturing efficiency,
a higher final energy, and improved beam quality preservation during the propagation.

This method of laser-based positron beam generation could be coupled with beam cooling methods to address the intrinsic high emittance of the bremsstrahlung-generated positron beam.  Such a system would make available a compact source of positron beams to test various plasma-based positron acceleration schemes. 

\begin{acknowledgments}
This work was supported by the Director, Office of Science, Office of High Energy Physics,
of the U.S. Department of Energy under Contract No. DE-AC02-05CH11231,
and used the computational facilities at the National Energy Research Scientific Computing Center (NERSC).
\end{acknowledgments}

\bibliographystyle{apsrev4-2}
\bibliography{positron_production_plasma_mirror}    

\begin{thebibliography}{46}%
\makeatletter
\providecommand \@ifxundefined [1]{%
 \@ifx{#1\undefined}
}%
\providecommand \@ifnum [1]{%
 \ifnum #1\expandafter \@firstoftwo
 \else \expandafter \@secondoftwo
 \fi
}%
\providecommand \@ifx [1]{%
 \ifx #1\expandafter \@firstoftwo
 \else \expandafter \@secondoftwo
 \fi
}%
\providecommand \natexlab [1]{#1}%
\providecommand \enquote  [1]{``#1''}%
\providecommand \bibnamefont  [1]{#1}%
\providecommand \bibfnamefont [1]{#1}%
\providecommand \citenamefont [1]{#1}%
\providecommand \href@noop [0]{\@secondoftwo}%
\providecommand \href [0]{\begingroup \@sanitize@url \@href}%
\providecommand \@href[1]{\@@startlink{#1}\@@href}%
\providecommand \@@href[1]{\endgroup#1\@@endlink}%
\providecommand \@sanitize@url [0]{\catcode `\\12\catcode `\$12\catcode `\&12\catcode `\#12\catcode `\^12\catcode `\_12\catcode `\%12\relax}%
\providecommand \@@startlink[1]{}%
\providecommand \@@endlink[0]{}%
\providecommand \url  [0]{\begingroup\@sanitize@url \@url }%
\providecommand \@url [1]{\endgroup\@href {#1}{\urlprefix }}%
\providecommand \urlprefix  [0]{URL }%
\providecommand \Eprint [0]{\href }%
\providecommand \doibase [0]{https://doi.org/}%
\providecommand \selectlanguage [0]{\@gobble}%
\providecommand \bibinfo  [0]{\@secondoftwo}%
\providecommand \bibfield  [0]{\@secondoftwo}%
\providecommand \translation [1]{[#1]}%
\providecommand \BibitemOpen [0]{}%
\providecommand \bibitemStop [0]{}%
\providecommand \bibitemNoStop [0]{.\EOS\space}%
\providecommand \EOS [0]{\spacefactor3000\relax}%
\providecommand \BibitemShut  [1]{\csname bibitem#1\endcsname}%
\let\auto@bib@innerbib\@empty
\bibitem [{\citenamefont {Esarey}\ \emph {et~al.}(2009)\citenamefont {Esarey}, \citenamefont {Schroeder},\ and\ \citenamefont {Leemans}}]{esarey_physics_2009}%
  \BibitemOpen
  \bibfield  {author} {\bibinfo {author} {\bibfnamefont {E.}~\bibnamefont {Esarey}}, \bibinfo {author} {\bibfnamefont {C.~B.}\ \bibnamefont {Schroeder}},\ and\ \bibinfo {author} {\bibfnamefont {W.~P.}\ \bibnamefont {Leemans}},\ }\href {https://doi.org/10.1103/RevModPhys.81.1229} {\bibfield  {journal} {\bibinfo  {journal} {Reviews of Modern Physics}\ }\textbf {\bibinfo {volume} {81}},\ \bibinfo {pages} {1229} (\bibinfo {year} {2009})},\ \bibinfo {note} {publisher: American Physical Society}\BibitemShut {NoStop}%
\bibitem [{\citenamefont {Hooker}(2013)}]{hooker_developments_2013}%
  \BibitemOpen
  \bibfield  {author} {\bibinfo {author} {\bibfnamefont {S.~M.}\ \bibnamefont {Hooker}},\ }\href {https://doi.org/10.1038/nphoton.2013.234} {\bibfield  {journal} {\bibinfo  {journal} {Nature Photonics}\ }\textbf {\bibinfo {volume} {7}},\ \bibinfo {pages} {775} (\bibinfo {year} {2013})},\ \bibinfo {note} {number: 10 Publisher: Nature Publishing Group}\BibitemShut {NoStop}%
\bibitem [{\citenamefont {Wang}\ \emph {et~al.}(2021)\citenamefont {Wang}, \citenamefont {Feng}, \citenamefont {Ke}, \citenamefont {Yu}, \citenamefont {Xu}, \citenamefont {Qi}, \citenamefont {Chen}, \citenamefont {Qin}, \citenamefont {Zhang}, \citenamefont {Fang}, \citenamefont {Liu}, \citenamefont {Jiang}, \citenamefont {Wang}, \citenamefont {Wang}, \citenamefont {Yang}, \citenamefont {Wu}, \citenamefont {Leng}, \citenamefont {Liu}, \citenamefont {Li},\ and\ \citenamefont {Xu}}]{wang_free-electron_2021}%
  \BibitemOpen
  \bibfield  {author} {\bibinfo {author} {\bibfnamefont {W.}~\bibnamefont {Wang}}, \bibinfo {author} {\bibfnamefont {K.}~\bibnamefont {Feng}}, \bibinfo {author} {\bibfnamefont {L.}~\bibnamefont {Ke}}, \bibinfo {author} {\bibfnamefont {C.}~\bibnamefont {Yu}}, \bibinfo {author} {\bibfnamefont {Y.}~\bibnamefont {Xu}}, \bibinfo {author} {\bibfnamefont {R.}~\bibnamefont {Qi}}, \bibinfo {author} {\bibfnamefont {Y.}~\bibnamefont {Chen}}, \bibinfo {author} {\bibfnamefont {Z.}~\bibnamefont {Qin}}, \bibinfo {author} {\bibfnamefont {Z.}~\bibnamefont {Zhang}}, \bibinfo {author} {\bibfnamefont {M.}~\bibnamefont {Fang}}, \bibinfo {author} {\bibfnamefont {J.}~\bibnamefont {Liu}}, \bibinfo {author} {\bibfnamefont {K.}~\bibnamefont {Jiang}}, \bibinfo {author} {\bibfnamefont {H.}~\bibnamefont {Wang}}, \bibinfo {author} {\bibfnamefont {C.}~\bibnamefont {Wang}}, \bibinfo {author} {\bibfnamefont {X.}~\bibnamefont {Yang}}, \bibinfo {author} {\bibfnamefont {F.}~\bibnamefont {Wu}}, \bibinfo {author} {\bibfnamefont {Y.}~\bibnamefont {Leng}}, \bibinfo {author} {\bibfnamefont {J.}~\bibnamefont {Liu}}, \bibinfo {author} {\bibfnamefont {R.}~\bibnamefont {Li}},\ and\ \bibinfo {author} {\bibfnamefont {Z.}~\bibnamefont {Xu}},\ }\href {https://doi.org/10.1038/s41586-021-03678-x} {\bibfield  {journal} {\bibinfo  {journal} {Nature}\ }\textbf {\bibinfo {volume} {595}},\ \bibinfo {pages} {516} (\bibinfo {year} {2021})},\ \bibinfo {note} {number: 7868 Publisher: Nature Publishing Group}\BibitemShut {NoStop}%
\bibitem [{\citenamefont {Labat}\ \emph {et~al.}(2023)\citenamefont {Labat}, \citenamefont {Cabada{\u g}}, \citenamefont {Ghaith}, \citenamefont {Irman}, \citenamefont {Berlioux}, \citenamefont {Berteaud}, \citenamefont {Blache}, \citenamefont {Bock}, \citenamefont {Bouvet}, \citenamefont {Briquez}, \citenamefont {Chang}, \citenamefont {Corde}, \citenamefont {Debus}, \citenamefont {De~Oliveira}, \citenamefont {Duval}, \citenamefont {Dietrich}, \citenamefont {El~Ajjouri}, \citenamefont {Eisenmann}, \citenamefont {Gautier}, \citenamefont {Gebhardt}, \citenamefont {Grams}, \citenamefont {Helbig}, \citenamefont {Herbeaux}, \citenamefont {Hubert}, \citenamefont {Kitegi}, \citenamefont {Kononenko}, \citenamefont {Kuntzsch}, \citenamefont {LaBerge}, \citenamefont {L{\^e}}, \citenamefont {Leluan}, \citenamefont {Loulergue}, \citenamefont {Malka}, \citenamefont {Marteau}, \citenamefont {Guyen}, \citenamefont {Oumbarek-Espinos}, \citenamefont {Pausch}, \citenamefont {Pereira}, \citenamefont {P{\"u}schel}, \citenamefont {Ricaud}, \citenamefont {Rommeluere}, \citenamefont {Roussel}, \citenamefont {Rousseau}, \citenamefont {Sch{\"o}bel}, \citenamefont {Sebdaoui}, \citenamefont {Steiniger}, \citenamefont {Tavakoli}, \citenamefont {Thaury}, \citenamefont {Ufer}, \citenamefont {Vall{\'e}au}, \citenamefont {Vandenberghe}, \citenamefont {V{\'e}t{\'e}ran}, \citenamefont {Schramm},\ and\ \citenamefont {Couprie}}]{labat_seeded_2023}%
  \BibitemOpen
  \bibfield  {author} {\bibinfo {author} {\bibfnamefont {M.}~\bibnamefont {Labat}}, \bibinfo {author} {\bibfnamefont {J.~C.}\ \bibnamefont {Cabada{\u g}}}, \bibinfo {author} {\bibfnamefont {A.}~\bibnamefont {Ghaith}}, \bibinfo {author} {\bibfnamefont {A.}~\bibnamefont {Irman}}, \bibinfo {author} {\bibfnamefont {A.}~\bibnamefont {Berlioux}}, \bibinfo {author} {\bibfnamefont {P.}~\bibnamefont {Berteaud}}, \bibinfo {author} {\bibfnamefont {F.}~\bibnamefont {Blache}}, \bibinfo {author} {\bibfnamefont {S.}~\bibnamefont {Bock}}, \bibinfo {author} {\bibfnamefont {F.}~\bibnamefont {Bouvet}}, \bibinfo {author} {\bibfnamefont {F.}~\bibnamefont {Briquez}}, \bibinfo {author} {\bibfnamefont {Y.-Y.}\ \bibnamefont {Chang}}, \bibinfo {author} {\bibfnamefont {S.}~\bibnamefont {Corde}}, \bibinfo {author} {\bibfnamefont {A.}~\bibnamefont {Debus}}, \bibinfo {author} {\bibfnamefont {C.}~\bibnamefont {De~Oliveira}}, \bibinfo {author} {\bibfnamefont {J.-P.}\ \bibnamefont {Duval}}, \bibinfo {author} {\bibfnamefont {Y.}~\bibnamefont {Dietrich}}, \bibinfo {author} {\bibfnamefont {M.}~\bibnamefont {El~Ajjouri}}, \bibinfo {author} {\bibfnamefont {C.}~\bibnamefont {Eisenmann}}, \bibinfo {author} {\bibfnamefont {J.}~\bibnamefont {Gautier}}, \bibinfo {author} {\bibfnamefont {R.}~\bibnamefont {Gebhardt}}, \bibinfo {author} {\bibfnamefont {S.}~\bibnamefont {Grams}}, \bibinfo {author} {\bibfnamefont {U.}~\bibnamefont {Helbig}}, \bibinfo {author} {\bibfnamefont {C.}~\bibnamefont {Herbeaux}}, \bibinfo {author} {\bibfnamefont {N.}~\bibnamefont {Hubert}}, \bibinfo {author} {\bibfnamefont {C.}~\bibnamefont {Kitegi}}, \bibinfo {author} {\bibfnamefont {O.}~\bibnamefont {Kononenko}}, \bibinfo {author} {\bibfnamefont {M.}~\bibnamefont {Kuntzsch}}, \bibinfo {author} {\bibfnamefont {M.}~\bibnamefont {LaBerge}}, \bibinfo {author} {\bibfnamefont {S.}~\bibnamefont {L{\^e}}}, \bibinfo {author} {\bibfnamefont {B.}~\bibnamefont {Leluan}}, \bibinfo {author} {\bibfnamefont {A.}~\bibnamefont {Loulergue}}, \bibinfo {author} {\bibfnamefont {V.}~\bibnamefont {Malka}}, \bibinfo {author} {\bibfnamefont {F.}~\bibnamefont {Marteau}}, \bibinfo {author} {\bibfnamefont {M.~H.~N.}\ \bibnamefont {Guyen}}, \bibinfo {author} {\bibfnamefont {D.}~\bibnamefont {Oumbarek-Espinos}}, \bibinfo {author} {\bibfnamefont {R.}~\bibnamefont {Pausch}}, \bibinfo {author} {\bibfnamefont {D.}~\bibnamefont {Pereira}}, \bibinfo {author} {\bibfnamefont {T.}~\bibnamefont {P{\"u}schel}}, \bibinfo {author} {\bibfnamefont {J.-P.}\ \bibnamefont {Ricaud}}, \bibinfo {author} {\bibfnamefont {P.}~\bibnamefont {Rommeluere}}, \bibinfo {author} {\bibfnamefont {E.}~\bibnamefont {Roussel}}, \bibinfo {author} {\bibfnamefont {P.}~\bibnamefont {Rousseau}}, \bibinfo {author} {\bibfnamefont {S.}~\bibnamefont {Sch{\"o}bel}}, \bibinfo {author} {\bibfnamefont {M.}~\bibnamefont {Sebdaoui}}, \bibinfo {author} {\bibfnamefont {K.}~\bibnamefont {Steiniger}}, \bibinfo {author} {\bibfnamefont {K.}~\bibnamefont {Tavakoli}}, \bibinfo {author} {\bibfnamefont {C.}~\bibnamefont {Thaury}}, \bibinfo {author} {\bibfnamefont {P.}~\bibnamefont {Ufer}}, \bibinfo {author} {\bibfnamefont {M.}~\bibnamefont {Vall{\'e}au}}, \bibinfo {author} {\bibfnamefont {M.}~\bibnamefont {Vandenberghe}}, \bibinfo {author} {\bibfnamefont {J.}~\bibnamefont {V{\'e}t{\'e}ran}}, \bibinfo {author} {\bibfnamefont {U.}~\bibnamefont {Schramm}},\ and\ \bibinfo {author} {\bibfnamefont {M.-E.}\ \bibnamefont {Couprie}},\ }\href {https://doi.org/10.1038/s41566-022-01104-w} {\bibfield  {journal} {\bibinfo  {journal} {Nature Photonics}\ }\textbf {\bibinfo {volume} {17}},\ \bibinfo {pages} {150} (\bibinfo {year} {2023})},\ \bibinfo {note} {number: 2 Publisher: Nature Publishing Group}\BibitemShut {NoStop}%
\bibitem [{\citenamefont {Benedetti}\ \emph {et~al.}(2022)\citenamefont {Benedetti}, \citenamefont {Bulanov}, \citenamefont {Esarey}, \citenamefont {Gonsalves}, \citenamefont {Jacobs}, \citenamefont {Knapen}, \citenamefont {Nachman}, \citenamefont {Nakamura}, \citenamefont {Griso}, \citenamefont {Schroeder}, \citenamefont {Terzani}, \citenamefont {van Tilborg}, \citenamefont {Turner}, \citenamefont {Yao}, \citenamefont {Bernstein}, \citenamefont {Shiltsev}, \citenamefont {Gessner}, \citenamefont {Hogan}, \citenamefont {Nelson}, \citenamefont {Jing}, \citenamefont {Low}, \citenamefont {Lu}, \citenamefont {Yoshida}, \citenamefont {Lee}, \citenamefont {Meade}, \citenamefont {Vafaei-Najafabadi}, \citenamefont {Muggli}, \citenamefont {Musumeci}, \citenamefont {Palmer}, \citenamefont {Prebys}, \citenamefont {Visinelli}, \citenamefont {Aidala},\ and\ \citenamefont {Thomas}}]{benedetti_whitepaper_2022}%
  \BibitemOpen
  \bibfield  {author} {\bibinfo {author} {\bibfnamefont {C.}~\bibnamefont {Benedetti}}, \bibinfo {author} {\bibfnamefont {S.~S.}\ \bibnamefont {Bulanov}}, \bibinfo {author} {\bibfnamefont {E.}~\bibnamefont {Esarey}}, \bibinfo {author} {\bibfnamefont {C.~G. R. G. A.~J.}\ \bibnamefont {Gonsalves}}, \bibinfo {author} {\bibfnamefont {P.~M.}\ \bibnamefont {Jacobs}}, \bibinfo {author} {\bibfnamefont {S.}~\bibnamefont {Knapen}}, \bibinfo {author} {\bibfnamefont {B.}~\bibnamefont {Nachman}}, \bibinfo {author} {\bibfnamefont {K.}~\bibnamefont {Nakamura}}, \bibinfo {author} {\bibfnamefont {S.~P.}\ \bibnamefont {Griso}}, \bibinfo {author} {\bibfnamefont {C.~B.}\ \bibnamefont {Schroeder}}, \bibinfo {author} {\bibfnamefont {D.}~\bibnamefont {Terzani}}, \bibinfo {author} {\bibfnamefont {J.}~\bibnamefont {van Tilborg}}, \bibinfo {author} {\bibfnamefont {M.}~\bibnamefont {Turner}}, \bibinfo {author} {\bibfnamefont {W.-M.}\ \bibnamefont {Yao}}, \bibinfo {author} {\bibfnamefont {R.}~\bibnamefont {Bernstein}}, \bibinfo {author} {\bibfnamefont {V.}~\bibnamefont {Shiltsev}}, \bibinfo {author} {\bibfnamefont {S.~J.}\ \bibnamefont {Gessner}}, \bibinfo {author} {\bibfnamefont {M.~J.}\ \bibnamefont {Hogan}}, \bibinfo {author} {\bibfnamefont {T.}~\bibnamefont {Nelson}}, \bibinfo {author} {\bibfnamefont {C.}~\bibnamefont {Jing}}, \bibinfo {author} {\bibfnamefont {I.}~\bibnamefont {Low}}, \bibinfo {author} {\bibfnamefont {X.}~\bibnamefont {Lu}}, \bibinfo {author} {\bibfnamefont {R.}~\bibnamefont {Yoshida}}, \bibinfo {author} {\bibfnamefont {C.}~\bibnamefont {Lee}}, \bibinfo {author} {\bibfnamefont {P.}~\bibnamefont {Meade}}, \bibinfo {author} {\bibfnamefont {N.}~\bibnamefont {Vafaei-Najafabadi}}, \bibinfo {author} {\bibfnamefont {P.}~\bibnamefont {Muggli}}, \bibinfo {author} {\bibfnamefont {P.}~\bibnamefont {Musumeci}}, \bibinfo {author} {\bibfnamefont {M.}~\bibnamefont {Palmer}}, \bibinfo {author} {\bibfnamefont {E.}~\bibnamefont {Prebys}}, \bibinfo {author} {\bibfnamefont {L.}~\bibnamefont {Visinelli}}, \bibinfo {author} {\bibfnamefont {C.~A.}\ \bibnamefont {Aidala}},\ and\ \bibinfo {author} {\bibfnamefont {A.~G.~R.}\ \bibnamefont {Thomas}},\ }\href {http://arxiv.org/abs/2203.08425} {\bibinfo {title} {Whitepaper submitted to {Snowmass21}: {Advanced} accelerator linear collider demonstration facility at intermediate energy}} (\bibinfo {year} {2022}),\ \bibinfo {note} {arXiv:2203.08425 [hep-ph, physics:physics]}\BibitemShut {NoStop}%
\bibitem [{\citenamefont {Schroeder}\ \emph {et~al.}(2023)\citenamefont {Schroeder}, \citenamefont {Albert}, \citenamefont {Benedetti}, \citenamefont {Bromage}, \citenamefont {Bruhwiler}, \citenamefont {Bulanov}, \citenamefont {Campbell}, \citenamefont {Cook}, \citenamefont {Cros}, \citenamefont {Downer}, \citenamefont {Esarey}, \citenamefont {Froula}, \citenamefont {Fuchs}, \citenamefont {Geddes}, \citenamefont {Gessner}, \citenamefont {Gonsalves}, \citenamefont {Hogan}, \citenamefont {Hooker}, \citenamefont {Huebl}, \citenamefont {Jing}, \citenamefont {Joshi}, \citenamefont {Krushelnick}, \citenamefont {Leemans}, \citenamefont {Lehe}, \citenamefont {Maier}, \citenamefont {Milchberg}, \citenamefont {Mori}, \citenamefont {Nakamura}, \citenamefont {Osterhoff}, \citenamefont {Palastro}, \citenamefont {Palmer}, \citenamefont {P{\~o}der}, \citenamefont {Power}, \citenamefont {Shadwick}, \citenamefont {Terzani}, \citenamefont {Th{\'e}venet}, \citenamefont {Thomas}, \citenamefont {van Tilborg}, \citenamefont {Turner}, \citenamefont {Vafaei-Najafabadi}, \citenamefont {Vay}, \citenamefont {Zhou},\ and\ \citenamefont {Zuegel}}]{benedetti_linear_2022}%
  \BibitemOpen
  \bibfield  {author} {\bibinfo {author} {\bibfnamefont {C.}~\bibnamefont {Schroeder}}, \bibinfo {author} {\bibfnamefont {F.}~\bibnamefont {Albert}}, \bibinfo {author} {\bibfnamefont {C.}~\bibnamefont {Benedetti}}, \bibinfo {author} {\bibfnamefont {J.}~\bibnamefont {Bromage}}, \bibinfo {author} {\bibfnamefont {D.}~\bibnamefont {Bruhwiler}}, \bibinfo {author} {\bibfnamefont {S.}~\bibnamefont {Bulanov}}, \bibinfo {author} {\bibfnamefont {E.}~\bibnamefont {Campbell}}, \bibinfo {author} {\bibfnamefont {N.}~\bibnamefont {Cook}}, \bibinfo {author} {\bibfnamefont {B.}~\bibnamefont {Cros}}, \bibinfo {author} {\bibfnamefont {M.}~\bibnamefont {Downer}}, \bibinfo {author} {\bibfnamefont {E.}~\bibnamefont {Esarey}}, \bibinfo {author} {\bibfnamefont {D.}~\bibnamefont {Froula}}, \bibinfo {author} {\bibfnamefont {M.}~\bibnamefont {Fuchs}}, \bibinfo {author} {\bibfnamefont {C.}~\bibnamefont {Geddes}}, \bibinfo {author} {\bibfnamefont {S.}~\bibnamefont {Gessner}}, \bibinfo {author} {\bibfnamefont {A.}~\bibnamefont {Gonsalves}}, \bibinfo {author} {\bibfnamefont {M.}~\bibnamefont {Hogan}}, \bibinfo {author} {\bibfnamefont {S.}~\bibnamefont {Hooker}}, \bibinfo {author} {\bibfnamefont {A.}~\bibnamefont {Huebl}}, \bibinfo {author} {\bibfnamefont {C.}~\bibnamefont {Jing}}, \bibinfo {author} {\bibfnamefont {C.}~\bibnamefont {Joshi}}, \bibinfo {author} {\bibfnamefont {K.}~\bibnamefont {Krushelnick}}, \bibinfo {author} {\bibfnamefont {W.}~\bibnamefont {Leemans}}, \bibinfo {author} {\bibfnamefont {R.}~\bibnamefont {Lehe}}, \bibinfo {author} {\bibfnamefont {A.}~\bibnamefont {Maier}}, \bibinfo {author} {\bibfnamefont {H.}~\bibnamefont {Milchberg}}, \bibinfo {author} {\bibfnamefont {W.}~\bibnamefont {Mori}}, \bibinfo {author} {\bibfnamefont {K.}~\bibnamefont {Nakamura}}, \bibinfo {author} {\bibfnamefont {J.}~\bibnamefont {Osterhoff}}, \bibinfo {author} {\bibfnamefont {J.}~\bibnamefont {Palastro}}, \bibinfo {author} {\bibfnamefont {M.}~\bibnamefont {Palmer}}, \bibinfo {author} {\bibfnamefont {K.}~\bibnamefont {P{\~o}der}}, \bibinfo {author} {\bibfnamefont {J.}~\bibnamefont {Power}}, \bibinfo {author} {\bibfnamefont {B.}~\bibnamefont {Shadwick}}, \bibinfo {author} {\bibfnamefont {D.}~\bibnamefont {Terzani}}, \bibinfo {author} {\bibfnamefont {M.}~\bibnamefont {Th{\'e}venet}}, \bibinfo {author} {\bibfnamefont {A.}~\bibnamefont {Thomas}}, \bibinfo {author} {\bibfnamefont {J.}~\bibnamefont {van Tilborg}}, \bibinfo {author} {\bibfnamefont {M.}~\bibnamefont {Turner}}, \bibinfo {author} {\bibfnamefont {N.}~\bibnamefont {Vafaei-Najafabadi}}, \bibinfo {author} {\bibfnamefont {J.-L.}\ \bibnamefont {Vay}}, \bibinfo {author} {\bibfnamefont {T.}~\bibnamefont {Zhou}},\ and\ \bibinfo {author} {\bibfnamefont {J.}~\bibnamefont {Zuegel}},\ }\href {https://doi.org/10.1088/1748-0221/18/06/T06001} {\bibfield  {journal} {\bibinfo  {journal} {Journal of Instrumentation}\ }\textbf {\bibinfo {volume} {18}}\bibinfo  {number} { (06)},\ \bibinfo {pages} {T06001}}\BibitemShut {NoStop}%
\bibitem [{\citenamefont {Faure}\ \emph {et~al.}(2004)\citenamefont {Faure}, \citenamefont {Glinec}, \citenamefont {Pukhov}, \citenamefont {Kiselev}, \citenamefont {Gordienko}, \citenamefont {Lefebvre}, \citenamefont {Rousseau}, \citenamefont {Burgy},\ and\ \citenamefont {Malka}}]{faure_laserplasma_2004}%
  \BibitemOpen
\bibfield  {number} {  }\bibfield  {author} {\bibinfo {author} {\bibfnamefont {J.}~\bibnamefont {Faure}}, \bibinfo {author} {\bibfnamefont {Y.}~\bibnamefont {Glinec}}, \bibinfo {author} {\bibfnamefont {A.}~\bibnamefont {Pukhov}}, \bibinfo {author} {\bibfnamefont {S.}~\bibnamefont {Kiselev}}, \bibinfo {author} {\bibfnamefont {S.}~\bibnamefont {Gordienko}}, \bibinfo {author} {\bibfnamefont {E.}~\bibnamefont {Lefebvre}}, \bibinfo {author} {\bibfnamefont {J.-P.}\ \bibnamefont {Rousseau}}, \bibinfo {author} {\bibfnamefont {F.}~\bibnamefont {Burgy}},\ and\ \bibinfo {author} {\bibfnamefont {V.}~\bibnamefont {Malka}},\ }\href {https://doi.org/10.1038/nature02963} {\bibfield  {journal} {\bibinfo  {journal} {Nature}\ }\textbf {\bibinfo {volume} {431}},\ \bibinfo {pages} {541} (\bibinfo {year} {2004})},\ \bibinfo {note} {number: 7008 Publisher: Nature Publishing Group}\BibitemShut {NoStop}%
\bibitem [{\citenamefont {Geddes}\ \emph {et~al.}(2004)\citenamefont {Geddes}, \citenamefont {Toth}, \citenamefont {van Tilborg}, \citenamefont {Esarey}, \citenamefont {Schroeder}, \citenamefont {Bruhwiler}, \citenamefont {Nieter}, \citenamefont {Cary},\ and\ \citenamefont {Leemans}}]{geddes_high-quality_2004}%
  \BibitemOpen
  \bibfield  {author} {\bibinfo {author} {\bibfnamefont {C.~G.~R.}\ \bibnamefont {Geddes}}, \bibinfo {author} {\bibfnamefont {C.}~\bibnamefont {Toth}}, \bibinfo {author} {\bibfnamefont {J.}~\bibnamefont {van Tilborg}}, \bibinfo {author} {\bibfnamefont {E.}~\bibnamefont {Esarey}}, \bibinfo {author} {\bibfnamefont {C.~B.}\ \bibnamefont {Schroeder}}, \bibinfo {author} {\bibfnamefont {D.}~\bibnamefont {Bruhwiler}}, \bibinfo {author} {\bibfnamefont {C.}~\bibnamefont {Nieter}}, \bibinfo {author} {\bibfnamefont {J.}~\bibnamefont {Cary}},\ and\ \bibinfo {author} {\bibfnamefont {W.~P.}\ \bibnamefont {Leemans}},\ }\href {https://doi.org/10.1038/nature02900} {\bibfield  {journal} {\bibinfo  {journal} {Nature}\ }\textbf {\bibinfo {volume} {431}},\ \bibinfo {pages} {538} (\bibinfo {year} {2004})},\ \bibinfo {note} {number: 7008 Publisher: Nature Publishing Group}\BibitemShut {NoStop}%
\bibitem [{\citenamefont {Mangles}\ \emph {et~al.}(2004)\citenamefont {Mangles}, \citenamefont {Murphy}, \citenamefont {Najmudin}, \citenamefont {Thomas}, \citenamefont {Collier}, \citenamefont {Dangor}, \citenamefont {Divall}, \citenamefont {Foster}, \citenamefont {Gallacher}, \citenamefont {Hooker}, \citenamefont {Jaroszynski}, \citenamefont {Langley}, \citenamefont {Mori}, \citenamefont {Norreys}, \citenamefont {Tsung}, \citenamefont {Viskup}, \citenamefont {Walton},\ and\ \citenamefont {Krushelnick}}]{mangles_monoenergetic_2004}%
  \BibitemOpen
  \bibfield  {author} {\bibinfo {author} {\bibfnamefont {S.~P.~D.}\ \bibnamefont {Mangles}}, \bibinfo {author} {\bibfnamefont {C.~D.}\ \bibnamefont {Murphy}}, \bibinfo {author} {\bibfnamefont {Z.}~\bibnamefont {Najmudin}}, \bibinfo {author} {\bibfnamefont {A.~G.~R.}\ \bibnamefont {Thomas}}, \bibinfo {author} {\bibfnamefont {J.~L.}\ \bibnamefont {Collier}}, \bibinfo {author} {\bibfnamefont {A.~E.}\ \bibnamefont {Dangor}}, \bibinfo {author} {\bibfnamefont {E.~J.}\ \bibnamefont {Divall}}, \bibinfo {author} {\bibfnamefont {P.~S.}\ \bibnamefont {Foster}}, \bibinfo {author} {\bibfnamefont {J.~G.}\ \bibnamefont {Gallacher}}, \bibinfo {author} {\bibfnamefont {C.~J.}\ \bibnamefont {Hooker}}, \bibinfo {author} {\bibfnamefont {D.~A.}\ \bibnamefont {Jaroszynski}}, \bibinfo {author} {\bibfnamefont {A.~J.}\ \bibnamefont {Langley}}, \bibinfo {author} {\bibfnamefont {W.~B.}\ \bibnamefont {Mori}}, \bibinfo {author} {\bibfnamefont {P.~A.}\ \bibnamefont {Norreys}}, \bibinfo {author} {\bibfnamefont {F.~S.}\ \bibnamefont {Tsung}}, \bibinfo {author} {\bibfnamefont {R.}~\bibnamefont {Viskup}}, \bibinfo {author} {\bibfnamefont {B.~R.}\ \bibnamefont {Walton}},\ and\ \bibinfo {author} {\bibfnamefont {K.}~\bibnamefont {Krushelnick}},\ }\href {https://doi.org/10.1038/nature02939} {\bibfield  {journal} {\bibinfo  {journal} {Nature}\ }\textbf {\bibinfo {volume} {431}},\ \bibinfo {pages} {535} (\bibinfo {year} {2004})},\ \bibinfo {note} {number: 7008 Publisher: Nature Publishing Group}\BibitemShut {NoStop}%
\bibitem [{\citenamefont {Wang}\ \emph {et~al.}(2013)\citenamefont {Wang}, \citenamefont {Zgadzaj}, \citenamefont {Fazel}, \citenamefont {Li}, \citenamefont {Yi}, \citenamefont {Zhang}, \citenamefont {Henderson}, \citenamefont {Chang}, \citenamefont {Korzekwa}, \citenamefont {Tsai}, \citenamefont {Pai}, \citenamefont {Quevedo}, \citenamefont {Dyer}, \citenamefont {Gaul}, \citenamefont {Martinez}, \citenamefont {Bernstein}, \citenamefont {Borger}, \citenamefont {Spinks}, \citenamefont {Donovan}, \citenamefont {Khudik}, \citenamefont {Shvets}, \citenamefont {Ditmire},\ and\ \citenamefont {Downer}}]{wang_quasi-monoenergetic_2013}%
  \BibitemOpen
  \bibfield  {author} {\bibinfo {author} {\bibfnamefont {X.}~\bibnamefont {Wang}}, \bibinfo {author} {\bibfnamefont {R.}~\bibnamefont {Zgadzaj}}, \bibinfo {author} {\bibfnamefont {N.}~\bibnamefont {Fazel}}, \bibinfo {author} {\bibfnamefont {Z.}~\bibnamefont {Li}}, \bibinfo {author} {\bibfnamefont {S.~A.}\ \bibnamefont {Yi}}, \bibinfo {author} {\bibfnamefont {X.}~\bibnamefont {Zhang}}, \bibinfo {author} {\bibfnamefont {W.}~\bibnamefont {Henderson}}, \bibinfo {author} {\bibfnamefont {Y.-Y.}\ \bibnamefont {Chang}}, \bibinfo {author} {\bibfnamefont {R.}~\bibnamefont {Korzekwa}}, \bibinfo {author} {\bibfnamefont {H.-E.}\ \bibnamefont {Tsai}}, \bibinfo {author} {\bibfnamefont {C.-H.}\ \bibnamefont {Pai}}, \bibinfo {author} {\bibfnamefont {H.}~\bibnamefont {Quevedo}}, \bibinfo {author} {\bibfnamefont {G.}~\bibnamefont {Dyer}}, \bibinfo {author} {\bibfnamefont {E.}~\bibnamefont {Gaul}}, \bibinfo {author} {\bibfnamefont {M.}~\bibnamefont {Martinez}}, \bibinfo {author} {\bibfnamefont {A.~C.}\ \bibnamefont {Bernstein}}, \bibinfo {author} {\bibfnamefont {T.}~\bibnamefont {Borger}}, \bibinfo {author} {\bibfnamefont {M.}~\bibnamefont {Spinks}}, \bibinfo {author} {\bibfnamefont {M.}~\bibnamefont {Donovan}}, \bibinfo {author} {\bibfnamefont {V.}~\bibnamefont {Khudik}}, \bibinfo {author} {\bibfnamefont {G.}~\bibnamefont {Shvets}}, \bibinfo {author} {\bibfnamefont {T.}~\bibnamefont {Ditmire}},\ and\ \bibinfo {author} {\bibfnamefont {M.~C.}\ \bibnamefont {Downer}},\ }\href {https://doi.org/10.1038/ncomms2988} {\bibfield  {journal} {\bibinfo  {journal} {Nature Communications}\ }\textbf {\bibinfo {volume} {4}},\ \bibinfo {pages} {1988} (\bibinfo {year} {2013})},\ \bibinfo {note} {number: 1 Publisher: Nature Publishing Group}\BibitemShut {NoStop}%
\bibitem [{\citenamefont {Gonsalves}\ \emph {et~al.}(2019)\citenamefont {Gonsalves}, \citenamefont {Nakamura}, \citenamefont {Daniels}, \citenamefont {Benedetti}, \citenamefont {Pieronek}, \citenamefont {de~Raadt}, \citenamefont {Steinke}, \citenamefont {Bin}, \citenamefont {Bulanov}, \citenamefont {van Tilborg}, \citenamefont {Geddes}, \citenamefont {Schroeder}, \citenamefont {T{\'o}th}, \citenamefont {Esarey}, \citenamefont {Swanson}, \citenamefont {Fan-Chiang}, \citenamefont {Bagdasarov}, \citenamefont {Bobrova}, \citenamefont {Gasilov}, \citenamefont {Korn}, \citenamefont {Sasorov},\ and\ \citenamefont {Leemans}}]{gonsalves_petawatt_2019}%
  \BibitemOpen
  \bibfield  {author} {\bibinfo {author} {\bibfnamefont {A.~J.}\ \bibnamefont {Gonsalves}}, \bibinfo {author} {\bibfnamefont {K.}~\bibnamefont {Nakamura}}, \bibinfo {author} {\bibfnamefont {J.}~\bibnamefont {Daniels}}, \bibinfo {author} {\bibfnamefont {C.}~\bibnamefont {Benedetti}}, \bibinfo {author} {\bibfnamefont {C.}~\bibnamefont {Pieronek}}, \bibinfo {author} {\bibfnamefont {T.~C.~H.}\ \bibnamefont {de~Raadt}}, \bibinfo {author} {\bibfnamefont {S.}~\bibnamefont {Steinke}}, \bibinfo {author} {\bibfnamefont {J.~H.}\ \bibnamefont {Bin}}, \bibinfo {author} {\bibfnamefont {S.~S.}\ \bibnamefont {Bulanov}}, \bibinfo {author} {\bibfnamefont {J.}~\bibnamefont {van Tilborg}}, \bibinfo {author} {\bibfnamefont {C.~G.~R.}\ \bibnamefont {Geddes}}, \bibinfo {author} {\bibfnamefont {C.~B.}\ \bibnamefont {Schroeder}}, \bibinfo {author} {\bibfnamefont {C.}~\bibnamefont {T{\'o}th}}, \bibinfo {author} {\bibfnamefont {E.}~\bibnamefont {Esarey}}, \bibinfo {author} {\bibfnamefont {K.}~\bibnamefont {Swanson}}, \bibinfo {author} {\bibfnamefont {L.}~\bibnamefont {Fan-Chiang}}, \bibinfo {author} {\bibfnamefont {G.}~\bibnamefont {Bagdasarov}}, \bibinfo {author} {\bibfnamefont {N.}~\bibnamefont {Bobrova}}, \bibinfo {author} {\bibfnamefont {V.}~\bibnamefont {Gasilov}}, \bibinfo {author} {\bibfnamefont {G.}~\bibnamefont {Korn}}, \bibinfo {author} {\bibfnamefont {P.}~\bibnamefont {Sasorov}},\ and\ \bibinfo {author} {\bibfnamefont {W.~P.}\ \bibnamefont {Leemans}},\ }\href {https://doi.org/10.1103/PhysRevLett.122.084801} {\bibfield  {journal} {\bibinfo  {journal} {Physical Review Letters}\ }\textbf {\bibinfo {volume} {122}},\ \bibinfo {pages} {084801} (\bibinfo {year} {2019})},\ \bibinfo {note} {publisher: American Physical Society}\BibitemShut {NoStop}%
\bibitem [{\citenamefont {Maier}\ \emph {et~al.}(2020)\citenamefont {Maier}, \citenamefont {Delbos}, \citenamefont {Eichner}, \citenamefont {H{\"u}bner}, \citenamefont {Jalas}, \citenamefont {Jeppe}, \citenamefont {Jolly}, \citenamefont {Kirchen}, \citenamefont {Leroux}, \citenamefont {Messner}, \citenamefont {Schnepp}, \citenamefont {Trunk}, \citenamefont {Walker}, \citenamefont {Werle},\ and\ \citenamefont {Winkler}}]{maier_decoding_2020}%
  \BibitemOpen
  \bibfield  {author} {\bibinfo {author} {\bibfnamefont {A.~R.}\ \bibnamefont {Maier}}, \bibinfo {author} {\bibfnamefont {N.~M.}\ \bibnamefont {Delbos}}, \bibinfo {author} {\bibfnamefont {T.}~\bibnamefont {Eichner}}, \bibinfo {author} {\bibfnamefont {L.}~\bibnamefont {H{\"u}bner}}, \bibinfo {author} {\bibfnamefont {S.}~\bibnamefont {Jalas}}, \bibinfo {author} {\bibfnamefont {L.}~\bibnamefont {Jeppe}}, \bibinfo {author} {\bibfnamefont {S.~W.}\ \bibnamefont {Jolly}}, \bibinfo {author} {\bibfnamefont {M.}~\bibnamefont {Kirchen}}, \bibinfo {author} {\bibfnamefont {V.}~\bibnamefont {Leroux}}, \bibinfo {author} {\bibfnamefont {P.}~\bibnamefont {Messner}}, \bibinfo {author} {\bibfnamefont {M.}~\bibnamefont {Schnepp}}, \bibinfo {author} {\bibfnamefont {M.}~\bibnamefont {Trunk}}, \bibinfo {author} {\bibfnamefont {P.~A.}\ \bibnamefont {Walker}}, \bibinfo {author} {\bibfnamefont {C.}~\bibnamefont {Werle}},\ and\ \bibinfo {author} {\bibfnamefont {P.}~\bibnamefont {Winkler}},\ }\href {https://doi.org/10.1103/PhysRevX.10.031039} {\bibfield  {journal} {\bibinfo  {journal} {Physical Review X}\ }\textbf {\bibinfo {volume} {10}},\ \bibinfo {pages} {031039} (\bibinfo {year} {2020})},\ \bibinfo {note} {publisher: American Physical Society}\BibitemShut {NoStop}%
\bibitem [{\citenamefont {Musumeci}\ \emph {et~al.}(2022)\citenamefont {Musumeci}, \citenamefont {Boffo}, \citenamefont {Bulanov}, \citenamefont {Chaikovska}, \citenamefont {Golfe}, \citenamefont {Gessner}, \citenamefont {Grames}, \citenamefont {Hessami}, \citenamefont {Ivanyushenkov}, \citenamefont {Lankford}, \citenamefont {Loisch}, \citenamefont {Moortgat-Pick}, \citenamefont {Nagaitsev}, \citenamefont {Riemann}, \citenamefont {Sievers}, \citenamefont {Tenholt},\ and\ \citenamefont {Yokoya}}]{musumeci_positron_2022}%
  \BibitemOpen
  \bibfield  {author} {\bibinfo {author} {\bibfnamefont {P.}~\bibnamefont {Musumeci}}, \bibinfo {author} {\bibfnamefont {C.}~\bibnamefont {Boffo}}, \bibinfo {author} {\bibfnamefont {S.~S.}\ \bibnamefont {Bulanov}}, \bibinfo {author} {\bibfnamefont {I.}~\bibnamefont {Chaikovska}}, \bibinfo {author} {\bibfnamefont {A.~F.}\ \bibnamefont {Golfe}}, \bibinfo {author} {\bibfnamefont {S.}~\bibnamefont {Gessner}}, \bibinfo {author} {\bibfnamefont {J.}~\bibnamefont {Grames}}, \bibinfo {author} {\bibfnamefont {R.}~\bibnamefont {Hessami}}, \bibinfo {author} {\bibfnamefont {Y.}~\bibnamefont {Ivanyushenkov}}, \bibinfo {author} {\bibfnamefont {A.}~\bibnamefont {Lankford}}, \bibinfo {author} {\bibfnamefont {G.}~\bibnamefont {Loisch}}, \bibinfo {author} {\bibfnamefont {G.}~\bibnamefont {Moortgat-Pick}}, \bibinfo {author} {\bibfnamefont {S.}~\bibnamefont {Nagaitsev}}, \bibinfo {author} {\bibfnamefont {S.}~\bibnamefont {Riemann}}, \bibinfo {author} {\bibfnamefont {P.}~\bibnamefont {Sievers}}, \bibinfo {author} {\bibfnamefont {C.}~\bibnamefont {Tenholt}},\ and\ \bibinfo {author} {\bibfnamefont {K.}~\bibnamefont {Yokoya}},\ }\href {http://arxiv.org/abs/2204.13245} {\bibinfo {title} {Positron {Sources} for {Future} {High} {Energy} {Physics} {Colliders}}} (\bibinfo {year} {2022}),\ \bibinfo {note} {arXiv:2204.13245 [physics]}\BibitemShut {NoStop}%
\bibitem [{\citenamefont {Hue}\ \emph {et~al.}(2021)\citenamefont {Hue}, \citenamefont {Cao}, \citenamefont {Andriyash}, \citenamefont {Knetsch}, \citenamefont {Hogan}, \citenamefont {Adli}, \citenamefont {Gessner},\ and\ \citenamefont {Corde}}]{hue_efficiency_2021}%
  \BibitemOpen
  \bibfield  {author} {\bibinfo {author} {\bibfnamefont {C.~S.}\ \bibnamefont {Hue}}, \bibinfo {author} {\bibfnamefont {G.~J.}\ \bibnamefont {Cao}}, \bibinfo {author} {\bibfnamefont {I.~A.}\ \bibnamefont {Andriyash}}, \bibinfo {author} {\bibfnamefont {A.}~\bibnamefont {Knetsch}}, \bibinfo {author} {\bibfnamefont {M.~J.}\ \bibnamefont {Hogan}}, \bibinfo {author} {\bibfnamefont {E.}~\bibnamefont {Adli}}, \bibinfo {author} {\bibfnamefont {S.}~\bibnamefont {Gessner}},\ and\ \bibinfo {author} {\bibfnamefont {S.}~\bibnamefont {Corde}},\ }\href {https://doi.org/10.1103/PhysRevResearch.3.043063} {\bibfield  {journal} {\bibinfo  {journal} {Physical Review Research}\ }\textbf {\bibinfo {volume} {3}},\ \bibinfo {pages} {043063} (\bibinfo {year} {2021})}\BibitemShut {NoStop}%
\bibitem [{\citenamefont {Silva}\ \emph {et~al.}(2021)\citenamefont {Silva}, \citenamefont {Amorim}, \citenamefont {Downer}, \citenamefont {Hogan}, \citenamefont {Yakimenko}, \citenamefont {Zgadzaj},\ and\ \citenamefont {Vieira}}]{silva_stable_2021}%
  \BibitemOpen
  \bibfield  {author} {\bibinfo {author} {\bibfnamefont {T.}~\bibnamefont {Silva}}, \bibinfo {author} {\bibfnamefont {L.~D.}\ \bibnamefont {Amorim}}, \bibinfo {author} {\bibfnamefont {M.~C.}\ \bibnamefont {Downer}}, \bibinfo {author} {\bibfnamefont {M.~J.}\ \bibnamefont {Hogan}}, \bibinfo {author} {\bibfnamefont {V.}~\bibnamefont {Yakimenko}}, \bibinfo {author} {\bibfnamefont {R.}~\bibnamefont {Zgadzaj}},\ and\ \bibinfo {author} {\bibfnamefont {J.}~\bibnamefont {Vieira}},\ }\href {https://doi.org/10.1103/PhysRevLett.127.104801} {\bibfield  {journal} {\bibinfo  {journal} {Physical Review Letters}\ }\textbf {\bibinfo {volume} {127}},\ \bibinfo {pages} {104801} (\bibinfo {year} {2021})},\ \bibinfo {note} {publisher: American Physical Society}\BibitemShut {NoStop}%
\bibitem [{\citenamefont {Zhou}\ \emph {et~al.}(2021)\citenamefont {Zhou}, \citenamefont {Hua}, \citenamefont {An}, \citenamefont {Mori}, \citenamefont {Joshi}, \citenamefont {Gao},\ and\ \citenamefont {Lu}}]{zhou_high_2021}%
  \BibitemOpen
  \bibfield  {author} {\bibinfo {author} {\bibfnamefont {S.}~\bibnamefont {Zhou}}, \bibinfo {author} {\bibfnamefont {J.}~\bibnamefont {Hua}}, \bibinfo {author} {\bibfnamefont {W.}~\bibnamefont {An}}, \bibinfo {author} {\bibfnamefont {W.~B.}\ \bibnamefont {Mori}}, \bibinfo {author} {\bibfnamefont {C.}~\bibnamefont {Joshi}}, \bibinfo {author} {\bibfnamefont {J.}~\bibnamefont {Gao}},\ and\ \bibinfo {author} {\bibfnamefont {W.}~\bibnamefont {Lu}},\ }\href {https://doi.org/10.1103/PhysRevLett.127.174801} {\bibfield  {journal} {\bibinfo  {journal} {Physical Review Letters}\ }\textbf {\bibinfo {volume} {127}},\ \bibinfo {pages} {174801} (\bibinfo {year} {2021})},\ \bibinfo {note} {publisher: American Physical Society}\BibitemShut {NoStop}%
\bibitem [{\citenamefont {Diederichs}\ \emph {et~al.}(2020)\citenamefont {Diederichs}, \citenamefont {Benedetti}, \citenamefont {Esarey}, \citenamefont {Osterhoff},\ and\ \citenamefont {Schroeder}}]{diederichs_high-quality_2020}%
  \BibitemOpen
  \bibfield  {author} {\bibinfo {author} {\bibfnamefont {S.}~\bibnamefont {Diederichs}}, \bibinfo {author} {\bibfnamefont {C.}~\bibnamefont {Benedetti}}, \bibinfo {author} {\bibfnamefont {E.}~\bibnamefont {Esarey}}, \bibinfo {author} {\bibfnamefont {J.}~\bibnamefont {Osterhoff}},\ and\ \bibinfo {author} {\bibfnamefont {C.~B.}\ \bibnamefont {Schroeder}},\ }\href {https://doi.org/10.1103/PhysRevAccelBeams.23.121301} {\bibfield  {journal} {\bibinfo  {journal} {Physical Review Accelerators and Beams}\ }\textbf {\bibinfo {volume} {23}},\ \bibinfo {pages} {121301} (\bibinfo {year} {2020})},\ \bibinfo {note} {publisher: American Physical Society}\BibitemShut {NoStop}%
\bibitem [{\citenamefont {Chen}\ \emph {et~al.}(2009)\citenamefont {Chen}, \citenamefont {Wilks}, \citenamefont {Bonlie}, \citenamefont {Liang}, \citenamefont {Myatt}, \citenamefont {Price}, \citenamefont {Meyerhofer},\ and\ \citenamefont {Beiersdorfer}}]{chen_relativistic_2009}%
  \BibitemOpen
  \bibfield  {author} {\bibinfo {author} {\bibfnamefont {H.}~\bibnamefont {Chen}}, \bibinfo {author} {\bibfnamefont {S.~C.}\ \bibnamefont {Wilks}}, \bibinfo {author} {\bibfnamefont {J.~D.}\ \bibnamefont {Bonlie}}, \bibinfo {author} {\bibfnamefont {E.~P.}\ \bibnamefont {Liang}}, \bibinfo {author} {\bibfnamefont {J.}~\bibnamefont {Myatt}}, \bibinfo {author} {\bibfnamefont {D.~F.}\ \bibnamefont {Price}}, \bibinfo {author} {\bibfnamefont {D.~D.}\ \bibnamefont {Meyerhofer}},\ and\ \bibinfo {author} {\bibfnamefont {P.}~\bibnamefont {Beiersdorfer}},\ }\href {https://doi.org/10.1103/PhysRevLett.102.105001} {\bibfield  {journal} {\bibinfo  {journal} {Physical Review Letters}\ }\textbf {\bibinfo {volume} {102}},\ \bibinfo {pages} {105001} (\bibinfo {year} {2009})},\ \bibinfo {note} {publisher: American Physical Society}\BibitemShut {NoStop}%
\bibitem [{\citenamefont {Chen}\ \emph {et~al.}(2010)\citenamefont {Chen}, \citenamefont {Wilks}, \citenamefont {Meyerhofer}, \citenamefont {Bonlie}, \citenamefont {Chen}, \citenamefont {Chen}, \citenamefont {Courtois}, \citenamefont {Elberson}, \citenamefont {Gregori}, \citenamefont {Kruer}, \citenamefont {Landoas}, \citenamefont {Mithen}, \citenamefont {Myatt}, \citenamefont {Murphy}, \citenamefont {Nilson}, \citenamefont {Price}, \citenamefont {Schneider}, \citenamefont {Shepherd}, \citenamefont {Stoeckl}, \citenamefont {Tabak}, \citenamefont {Tommasini},\ and\ \citenamefont {Beiersdorfer}}]{chen_relativistic_2010}%
  \BibitemOpen
  \bibfield  {author} {\bibinfo {author} {\bibfnamefont {H.}~\bibnamefont {Chen}}, \bibinfo {author} {\bibfnamefont {S.~C.}\ \bibnamefont {Wilks}}, \bibinfo {author} {\bibfnamefont {D.~D.}\ \bibnamefont {Meyerhofer}}, \bibinfo {author} {\bibfnamefont {J.}~\bibnamefont {Bonlie}}, \bibinfo {author} {\bibfnamefont {C.~D.}\ \bibnamefont {Chen}}, \bibinfo {author} {\bibfnamefont {S.~N.}\ \bibnamefont {Chen}}, \bibinfo {author} {\bibfnamefont {C.}~\bibnamefont {Courtois}}, \bibinfo {author} {\bibfnamefont {L.}~\bibnamefont {Elberson}}, \bibinfo {author} {\bibfnamefont {G.}~\bibnamefont {Gregori}}, \bibinfo {author} {\bibfnamefont {W.}~\bibnamefont {Kruer}}, \bibinfo {author} {\bibfnamefont {O.}~\bibnamefont {Landoas}}, \bibinfo {author} {\bibfnamefont {J.}~\bibnamefont {Mithen}}, \bibinfo {author} {\bibfnamefont {J.}~\bibnamefont {Myatt}}, \bibinfo {author} {\bibfnamefont {C.~D.}\ \bibnamefont {Murphy}}, \bibinfo {author} {\bibfnamefont {P.}~\bibnamefont {Nilson}}, \bibinfo {author} {\bibfnamefont {D.}~\bibnamefont {Price}}, \bibinfo {author} {\bibfnamefont {M.}~\bibnamefont {Schneider}}, \bibinfo {author} {\bibfnamefont {R.}~\bibnamefont {Shepherd}}, \bibinfo {author} {\bibfnamefont {C.}~\bibnamefont {Stoeckl}}, \bibinfo {author} {\bibfnamefont {M.}~\bibnamefont {Tabak}}, \bibinfo {author} {\bibfnamefont {R.}~\bibnamefont {Tommasini}},\ and\ \bibinfo {author} {\bibfnamefont {P.}~\bibnamefont {Beiersdorfer}},\ }\href {https://doi.org/10.1103/PhysRevLett.105.015003} {\bibfield  {journal} {\bibinfo  {journal} {Physical Review Letters}\ }\textbf {\bibinfo {volume} {105}},\ \bibinfo {pages} {015003} (\bibinfo {year} {2010})},\ \bibinfo {note} {publisher: American Physical Society}\BibitemShut {NoStop}%
\bibitem [{\citenamefont {Sarri}\ \emph {et~al.}(2013)\citenamefont {Sarri}, \citenamefont {Schumaker}, \citenamefont {Di~Piazza}, \citenamefont {Poder}, \citenamefont {Cole}, \citenamefont {Vargas}, \citenamefont {Doria}, \citenamefont {Kushel}, \citenamefont {Dromey}, \citenamefont {Grittani}, \citenamefont {Gizzi}, \citenamefont {Dieckmann}, \citenamefont {Green}, \citenamefont {Chvykov}, \citenamefont {Maksimchuk}, \citenamefont {Yanovsky}, \citenamefont {He}, \citenamefont {Hou}, \citenamefont {Nees}, \citenamefont {Kar}, \citenamefont {Najmudin}, \citenamefont {Thomas}, \citenamefont {Keitel}, \citenamefont {Krushelnick},\ and\ \citenamefont {Zepf}}]{sarri_laser-driven_2013}%
  \BibitemOpen
  \bibfield  {author} {\bibinfo {author} {\bibfnamefont {G.}~\bibnamefont {Sarri}}, \bibinfo {author} {\bibfnamefont {W.}~\bibnamefont {Schumaker}}, \bibinfo {author} {\bibfnamefont {A.}~\bibnamefont {Di~Piazza}}, \bibinfo {author} {\bibfnamefont {K.}~\bibnamefont {Poder}}, \bibinfo {author} {\bibfnamefont {J.~M.}\ \bibnamefont {Cole}}, \bibinfo {author} {\bibfnamefont {M.}~\bibnamefont {Vargas}}, \bibinfo {author} {\bibfnamefont {D.}~\bibnamefont {Doria}}, \bibinfo {author} {\bibfnamefont {S.}~\bibnamefont {Kushel}}, \bibinfo {author} {\bibfnamefont {B.}~\bibnamefont {Dromey}}, \bibinfo {author} {\bibfnamefont {G.}~\bibnamefont {Grittani}}, \bibinfo {author} {\bibfnamefont {L.}~\bibnamefont {Gizzi}}, \bibinfo {author} {\bibfnamefont {M.~E.}\ \bibnamefont {Dieckmann}}, \bibinfo {author} {\bibfnamefont {A.}~\bibnamefont {Green}}, \bibinfo {author} {\bibfnamefont {V.}~\bibnamefont {Chvykov}}, \bibinfo {author} {\bibfnamefont {A.}~\bibnamefont {Maksimchuk}}, \bibinfo {author} {\bibfnamefont {V.}~\bibnamefont {Yanovsky}}, \bibinfo {author} {\bibfnamefont {Z.~H.}\ \bibnamefont {He}}, \bibinfo {author} {\bibfnamefont {B.~X.}\ \bibnamefont {Hou}}, \bibinfo {author} {\bibfnamefont {J.~A.}\ \bibnamefont {Nees}}, \bibinfo {author} {\bibfnamefont {S.}~\bibnamefont {Kar}}, \bibinfo {author} {\bibfnamefont {Z.}~\bibnamefont {Najmudin}}, \bibinfo {author} {\bibfnamefont {A.~G.~R.}\ \bibnamefont {Thomas}}, \bibinfo {author} {\bibfnamefont {C.~H.}\ \bibnamefont {Keitel}}, \bibinfo {author} {\bibfnamefont {K.}~\bibnamefont {Krushelnick}},\ and\ \bibinfo {author} {\bibfnamefont {M.}~\bibnamefont {Zepf}},\ }\href {https://doi.org/10.1088/0741-3335/55/12/124017} {\bibfield  {journal} {\bibinfo  {journal} {Plasma Physics and Controlled Fusion}\ }\textbf {\bibinfo {volume} {55}},\ \bibinfo {pages} {124017} (\bibinfo {year} {2013})}\BibitemShut {NoStop}%
\bibitem [{\citenamefont {Sarri}\ \emph {et~al.}(2015)\citenamefont {Sarri}, \citenamefont {Poder}, \citenamefont {Cole}, \citenamefont {Schumaker}, \citenamefont {Di~Piazza}, \citenamefont {Reville}, \citenamefont {Dzelzainis}, \citenamefont {Doria}, \citenamefont {Gizzi}, \citenamefont {Grittani}, \citenamefont {Kar}, \citenamefont {Keitel}, \citenamefont {Krushelnick}, \citenamefont {Kuschel}, \citenamefont {Mangles}, \citenamefont {Najmudin}, \citenamefont {Shukla}, \citenamefont {Silva}, \citenamefont {Symes}, \citenamefont {Thomas}, \citenamefont {Vargas}, \citenamefont {Vieira},\ and\ \citenamefont {Zepf}}]{sarri_generation_2015}%
  \BibitemOpen
  \bibfield  {author} {\bibinfo {author} {\bibfnamefont {G.}~\bibnamefont {Sarri}}, \bibinfo {author} {\bibfnamefont {K.}~\bibnamefont {Poder}}, \bibinfo {author} {\bibfnamefont {J.~M.}\ \bibnamefont {Cole}}, \bibinfo {author} {\bibfnamefont {W.}~\bibnamefont {Schumaker}}, \bibinfo {author} {\bibfnamefont {A.}~\bibnamefont {Di~Piazza}}, \bibinfo {author} {\bibfnamefont {B.}~\bibnamefont {Reville}}, \bibinfo {author} {\bibfnamefont {T.}~\bibnamefont {Dzelzainis}}, \bibinfo {author} {\bibfnamefont {D.}~\bibnamefont {Doria}}, \bibinfo {author} {\bibfnamefont {L.~A.}\ \bibnamefont {Gizzi}}, \bibinfo {author} {\bibfnamefont {G.}~\bibnamefont {Grittani}}, \bibinfo {author} {\bibfnamefont {S.}~\bibnamefont {Kar}}, \bibinfo {author} {\bibfnamefont {C.~H.}\ \bibnamefont {Keitel}}, \bibinfo {author} {\bibfnamefont {K.}~\bibnamefont {Krushelnick}}, \bibinfo {author} {\bibfnamefont {S.}~\bibnamefont {Kuschel}}, \bibinfo {author} {\bibfnamefont {S.~P.~D.}\ \bibnamefont {Mangles}}, \bibinfo {author} {\bibfnamefont {Z.}~\bibnamefont {Najmudin}}, \bibinfo {author} {\bibfnamefont {N.}~\bibnamefont {Shukla}}, \bibinfo {author} {\bibfnamefont {L.~O.}\ \bibnamefont {Silva}}, \bibinfo {author} {\bibfnamefont {D.}~\bibnamefont {Symes}}, \bibinfo {author} {\bibfnamefont {A.~G.~R.}\ \bibnamefont {Thomas}}, \bibinfo {author} {\bibfnamefont {M.}~\bibnamefont {Vargas}}, \bibinfo {author} {\bibfnamefont {J.}~\bibnamefont {Vieira}},\ and\ \bibinfo {author} {\bibfnamefont {M.}~\bibnamefont {Zepf}},\ }\href {https://doi.org/10.1038/ncomms7747} {\bibfield  {journal} {\bibinfo  {journal} {Nature Communications}\ }\textbf {\bibinfo {volume} {6}},\ \bibinfo {pages} {6747} (\bibinfo {year} {2015})},\ \bibinfo {note} {number: 1 Publisher: Nature Publishing Group}\BibitemShut {NoStop}%
\bibitem [{\citenamefont {Fujii}\ \emph {et~al.}(2019)\citenamefont {Fujii}, \citenamefont {Marsh}, \citenamefont {An}, \citenamefont {Corde}, \citenamefont {Hogan}, \citenamefont {Yakimenko},\ and\ \citenamefont {Joshi}}]{fujii_positron_2019}%
  \BibitemOpen
  \bibfield  {author} {\bibinfo {author} {\bibfnamefont {H.}~\bibnamefont {Fujii}}, \bibinfo {author} {\bibfnamefont {K.~A.}\ \bibnamefont {Marsh}}, \bibinfo {author} {\bibfnamefont {W.}~\bibnamefont {An}}, \bibinfo {author} {\bibfnamefont {S.}~\bibnamefont {Corde}}, \bibinfo {author} {\bibfnamefont {M.~J.}\ \bibnamefont {Hogan}}, \bibinfo {author} {\bibfnamefont {V.}~\bibnamefont {Yakimenko}},\ and\ \bibinfo {author} {\bibfnamefont {C.}~\bibnamefont {Joshi}},\ }\href {https://doi.org/10.1103/PhysRevAccelBeams.22.091301} {\bibfield  {journal} {\bibinfo  {journal} {Physical Review Accelerators and Beams}\ }\textbf {\bibinfo {volume} {22}},\ \bibinfo {pages} {091301} (\bibinfo {year} {2019})},\ \bibinfo {note} {publisher: American Physical Society}\BibitemShut {NoStop}%
\bibitem [{\citenamefont {Streeter}\ \emph {et~al.}(2022)\citenamefont {Streeter}, \citenamefont {Colgan}, \citenamefont {Cavanagh}, \citenamefont {Los}, \citenamefont {Antoine}, \citenamefont {Audet}, \citenamefont {Balcazar}, \citenamefont {Calvin}, \citenamefont {Cardarelli}, \citenamefont {Ahmed}, \citenamefont {Kettle}, \citenamefont {Ma}, \citenamefont {Mangles}, \citenamefont {Najmudin}, \citenamefont {Rajeev}, \citenamefont {Symes}, \citenamefont {Thomas},\ and\ \citenamefont {Sarri}}]{streeter_narrow_2022}%
  \BibitemOpen
  \bibfield  {author} {\bibinfo {author} {\bibfnamefont {M.}~\bibnamefont {Streeter}}, \bibinfo {author} {\bibfnamefont {C.}~\bibnamefont {Colgan}}, \bibinfo {author} {\bibfnamefont {N.}~\bibnamefont {Cavanagh}}, \bibinfo {author} {\bibfnamefont {E.}~\bibnamefont {Los}}, \bibinfo {author} {\bibfnamefont {A.}~\bibnamefont {Antoine}}, \bibinfo {author} {\bibfnamefont {T.}~\bibnamefont {Audet}}, \bibinfo {author} {\bibfnamefont {M.}~\bibnamefont {Balcazar}}, \bibinfo {author} {\bibfnamefont {L.}~\bibnamefont {Calvin}}, \bibinfo {author} {\bibfnamefont {J.}~\bibnamefont {Cardarelli}}, \bibinfo {author} {\bibfnamefont {H.}~\bibnamefont {Ahmed}}, \bibinfo {author} {\bibfnamefont {B.}~\bibnamefont {Kettle}}, \bibinfo {author} {\bibfnamefont {Y.}~\bibnamefont {Ma}}, \bibinfo {author} {\bibfnamefont {S.}~\bibnamefont {Mangles}}, \bibinfo {author} {\bibfnamefont {Z.}~\bibnamefont {Najmudin}}, \bibinfo {author} {\bibfnamefont {P.~P.}\ \bibnamefont {Rajeev}}, \bibinfo {author} {\bibfnamefont {D.}~\bibnamefont {Symes}}, \bibinfo {author} {\bibfnamefont {A.}~\bibnamefont {Thomas}},\ and\ \bibinfo {author} {\bibfnamefont {G.}~\bibnamefont {Sarri}},\ }\href {https://doi.org/10.21203/rs.3.rs-2191274/v1} {\emph {\bibinfo {title} {Narrow bandwidth, low-emittance positron beams from a laser-wakefield accelerator}}},\ \bibinfo {type} {preprint}\ (\bibinfo  {institution} {In Review},\ \bibinfo {year} {2022})\BibitemShut {NoStop}%
\bibitem [{\citenamefont {Lobet}\ \emph {et~al.}(2017)\citenamefont {Lobet}, \citenamefont {Davoine}, \citenamefont {d’Humières},\ and\ \citenamefont {Gremillet}}]{lobet_generation_2017}%
  \BibitemOpen
  \bibfield  {author} {\bibinfo {author} {\bibfnamefont {M.}~\bibnamefont {Lobet}}, \bibinfo {author} {\bibfnamefont {X.}~\bibnamefont {Davoine}}, \bibinfo {author} {\bibfnamefont {E.}~\bibnamefont {d’Humières}},\ and\ \bibinfo {author} {\bibfnamefont {L.}~\bibnamefont {Gremillet}},\ }\href {https://doi.org/10.1103/PhysRevAccelBeams.20.043401} {\bibfield  {journal} {\bibinfo  {journal} {Physical Review Accelerators and Beams}\ }\textbf {\bibinfo {volume} {20}},\ \bibinfo {pages} {043401} (\bibinfo {year} {2017})}\BibitemShut {NoStop}%
\bibitem [{\citenamefont {Vranic}\ \emph {et~al.}(2018)\citenamefont {Vranic}, \citenamefont {Klimo}, \citenamefont {Korn},\ and\ \citenamefont {Weber}}]{vranic_multi-gev_2018}%
  \BibitemOpen
  \bibfield  {author} {\bibinfo {author} {\bibfnamefont {M.}~\bibnamefont {Vranic}}, \bibinfo {author} {\bibfnamefont {O.}~\bibnamefont {Klimo}}, \bibinfo {author} {\bibfnamefont {G.}~\bibnamefont {Korn}},\ and\ \bibinfo {author} {\bibfnamefont {S.}~\bibnamefont {Weber}},\ }\href {https://doi.org/10.1038/s41598-018-23126-7} {\bibfield  {journal} {\bibinfo  {journal} {Scientific Reports}\ }\textbf {\bibinfo {volume} {8}},\ \bibinfo {pages} {4702} (\bibinfo {year} {2018})}\BibitemShut {NoStop}%
\bibitem [{\citenamefont {Gonoskov}\ \emph {et~al.}(2022)\citenamefont {Gonoskov}, \citenamefont {Blackburn}, \citenamefont {Marklund},\ and\ \citenamefont {Bulanov}}]{gonoskov_charged_2022}%
  \BibitemOpen
  \bibfield  {author} {\bibinfo {author} {\bibfnamefont {A.}~\bibnamefont {Gonoskov}}, \bibinfo {author} {\bibfnamefont {T.~G.}\ \bibnamefont {Blackburn}}, \bibinfo {author} {\bibfnamefont {M.}~\bibnamefont {Marklund}},\ and\ \bibinfo {author} {\bibfnamefont {S.~S.}\ \bibnamefont {Bulanov}},\ }\href {https://doi.org/10.1103/RevModPhys.94.045001} {\bibfield  {journal} {\bibinfo  {journal} {Reviews of Modern Physics}\ }\textbf {\bibinfo {volume} {94}},\ \bibinfo {pages} {045001} (\bibinfo {year} {2022})},\ \bibinfo {note} {publisher: American Physical Society}\BibitemShut {NoStop}%
\bibitem [{\citenamefont {Zhu}\ \emph {et~al.}(2016)\citenamefont {Zhu}, \citenamefont {Yu}, \citenamefont {Sheng}, \citenamefont {Yin}, \citenamefont {Turcu},\ and\ \citenamefont {Pukhov}}]{zhu_dense_2016}%
  \BibitemOpen
  \bibfield  {author} {\bibinfo {author} {\bibfnamefont {X.-L.}\ \bibnamefont {Zhu}}, \bibinfo {author} {\bibfnamefont {T.-P.}\ \bibnamefont {Yu}}, \bibinfo {author} {\bibfnamefont {Z.-M.}\ \bibnamefont {Sheng}}, \bibinfo {author} {\bibfnamefont {Y.}~\bibnamefont {Yin}}, \bibinfo {author} {\bibfnamefont {I.~C.~E.}\ \bibnamefont {Turcu}},\ and\ \bibinfo {author} {\bibfnamefont {A.}~\bibnamefont {Pukhov}},\ }\href {https://doi.org/10.1038/ncomms13686} {\bibfield  {journal} {\bibinfo  {journal} {Nature Communications}\ }\textbf {\bibinfo {volume} {7}},\ \bibinfo {pages} {13686} (\bibinfo {year} {2016})}\BibitemShut {NoStop}%
\bibitem [{\citenamefont {Zhao}\ \emph {et~al.}(2022)\citenamefont {Zhao}, \citenamefont {Hu}, \citenamefont {Lu}, \citenamefont {Zhang}, \citenamefont {Hu}, \citenamefont {Zhu}, \citenamefont {Sheng}, \citenamefont {Turcu}, \citenamefont {Pukhov}, \citenamefont {Shao},\ and\ \citenamefont {Yu}}]{zhao_all-optical_2022}%
  \BibitemOpen
  \bibfield  {author} {\bibinfo {author} {\bibfnamefont {J.}~\bibnamefont {Zhao}}, \bibinfo {author} {\bibfnamefont {Y.-T.}\ \bibnamefont {Hu}}, \bibinfo {author} {\bibfnamefont {Y.}~\bibnamefont {Lu}}, \bibinfo {author} {\bibfnamefont {H.}~\bibnamefont {Zhang}}, \bibinfo {author} {\bibfnamefont {L.-X.}\ \bibnamefont {Hu}}, \bibinfo {author} {\bibfnamefont {X.-L.}\ \bibnamefont {Zhu}}, \bibinfo {author} {\bibfnamefont {Z.-M.}\ \bibnamefont {Sheng}}, \bibinfo {author} {\bibfnamefont {I.~C.~E.}\ \bibnamefont {Turcu}}, \bibinfo {author} {\bibfnamefont {A.}~\bibnamefont {Pukhov}}, \bibinfo {author} {\bibfnamefont {F.-Q.}\ \bibnamefont {Shao}},\ and\ \bibinfo {author} {\bibfnamefont {T.-P.}\ \bibnamefont {Yu}},\ }\href {https://doi.org/10.1038/s42005-021-00797-9} {\bibfield  {journal} {\bibinfo  {journal} {Communications Physics}\ }\textbf {\bibinfo {volume} {5}},\ \bibinfo {pages} {1} (\bibinfo {year} {2022})}\BibitemShut {NoStop}%
\bibitem [{\citenamefont {Martinez}\ \emph {et~al.}(2023)\citenamefont {Martinez}, \citenamefont {Barbosa},\ and\ \citenamefont {Vranic}}]{martinez_creation_2023}%
  \BibitemOpen
  \bibfield  {author} {\bibinfo {author} {\bibfnamefont {B.}~\bibnamefont {Martinez}}, \bibinfo {author} {\bibfnamefont {B.}~\bibnamefont {Barbosa}},\ and\ \bibinfo {author} {\bibfnamefont {M.}~\bibnamefont {Vranic}},\ }\href {https://doi.org/10.1103/PhysRevAccelBeams.26.011301} {\bibfield  {journal} {\bibinfo  {journal} {Physical Review Accelerators and Beams}\ }\textbf {\bibinfo {volume} {26}},\ \bibinfo {pages} {011301} (\bibinfo {year} {2023})},\ \bibinfo {note} {publisher: American Physical Society}\BibitemShut {NoStop}%
\bibitem [{\citenamefont {Chao}\ \emph {et~al.}(2013)\citenamefont {Chao}, \citenamefont {Mess},\ and\ \citenamefont {{others}}}]{chao_handbook_2013}%
  \BibitemOpen
  \bibfield  {author} {\bibinfo {author} {\bibfnamefont {A.~W.}\ \bibnamefont {Chao}}, \bibinfo {author} {\bibfnamefont {K.~H.}\ \bibnamefont {Mess}},\ and\ \bibinfo {author} {\bibnamefont {{others}}},\ }\href@noop {} {\emph {\bibinfo {title} {Handbook of accelerator physics and engineering}}}\ (\bibinfo  {publisher} {World scientific},\ \bibinfo {year} {2013})\BibitemShut {NoStop}%
\bibitem [{\citenamefont {Sokollik}\ \emph {et~al.}(2010)\citenamefont {Sokollik}, \citenamefont {Shiraishi}, \citenamefont {Osterhoff}, \citenamefont {Evans}, \citenamefont {Gonsalves}, \citenamefont {Nakamura}, \citenamefont {van Tilborg}, \citenamefont {Lin}, \citenamefont {Toth},\ and\ \citenamefont {Leemans}}]{sokollik_tapedrive_2010}%
  \BibitemOpen
  \bibfield  {author} {\bibinfo {author} {\bibfnamefont {T.}~\bibnamefont {Sokollik}}, \bibinfo {author} {\bibfnamefont {S.}~\bibnamefont {Shiraishi}}, \bibinfo {author} {\bibfnamefont {J.}~\bibnamefont {Osterhoff}}, \bibinfo {author} {\bibfnamefont {E.}~\bibnamefont {Evans}}, \bibinfo {author} {\bibfnamefont {A.~J.}\ \bibnamefont {Gonsalves}}, \bibinfo {author} {\bibfnamefont {K.}~\bibnamefont {Nakamura}}, \bibinfo {author} {\bibfnamefont {J.}~\bibnamefont {van Tilborg}}, \bibinfo {author} {\bibfnamefont {C.}~\bibnamefont {Lin}}, \bibinfo {author} {\bibfnamefont {C.}~\bibnamefont {Toth}},\ and\ \bibinfo {author} {\bibfnamefont {W.~P.}\ \bibnamefont {Leemans}},\ }\href {https://doi.org/10.1063/1.3520320} {\bibfield  {journal} {\bibinfo  {journal} {AIP Conference Proceedings}\ }\textbf {\bibinfo {volume} {1299}},\ \bibinfo {pages} {233} (\bibinfo {year} {2010})},\ \bibinfo {note} {publisher: American Institute of Physics}\BibitemShut {NoStop}%
\bibitem [{\citenamefont {Shaw}\ \emph {et~al.}(2016)\citenamefont {Shaw}, \citenamefont {Steinke}, \citenamefont {van Tilborg},\ and\ \citenamefont {Leemans}}]{shaw_reflectance_2016}%
  \BibitemOpen
  \bibfield  {author} {\bibinfo {author} {\bibfnamefont {B.~H.}\ \bibnamefont {Shaw}}, \bibinfo {author} {\bibfnamefont {S.}~\bibnamefont {Steinke}}, \bibinfo {author} {\bibfnamefont {J.}~\bibnamefont {van Tilborg}},\ and\ \bibinfo {author} {\bibfnamefont {W.~P.}\ \bibnamefont {Leemans}},\ }\href {https://doi.org/10.1063/1.4954242} {\bibfield  {journal} {\bibinfo  {journal} {Physics of Plasmas}\ }\textbf {\bibinfo {volume} {23}},\ \bibinfo {pages} {063118} (\bibinfo {year} {2016})},\ \bibinfo {note} {publisher: American Institute of Physics}\BibitemShut {NoStop}%
\bibitem [{\citenamefont {Amorim}\ \emph {et~al.}(2023)\citenamefont {Amorim}, \citenamefont {Benedetti}, \citenamefont {Bulanov}, \citenamefont {Terzani}, \citenamefont {Huebl}, \citenamefont {Schroeder}, \citenamefont {Vay},\ and\ \citenamefont {Esarey}}]{amorim_design_2023}%
  \BibitemOpen
  \bibfield  {author} {\bibinfo {author} {\bibfnamefont {L.~D.}\ \bibnamefont {Amorim}}, \bibinfo {author} {\bibfnamefont {C.}~\bibnamefont {Benedetti}}, \bibinfo {author} {\bibfnamefont {S.~S.}\ \bibnamefont {Bulanov}}, \bibinfo {author} {\bibfnamefont {D.}~\bibnamefont {Terzani}}, \bibinfo {author} {\bibfnamefont {A.}~\bibnamefont {Huebl}}, \bibinfo {author} {\bibfnamefont {C.~B.}\ \bibnamefont {Schroeder}}, \bibinfo {author} {\bibfnamefont {J.-L.}\ \bibnamefont {Vay}},\ and\ \bibinfo {author} {\bibfnamefont {E.}~\bibnamefont {Esarey}},\ }\bibfield  {journal} {\bibinfo  {journal} {Plasma Physics and Controlled Fusion}\ }\href {https://doi.org/10.1088/1361-6587/ace3f1} {10.1088/1361-6587/ace3f1} (\bibinfo {year} {2023})\BibitemShut {NoStop}%
\bibitem [{\citenamefont {Sahai}(2018)}]{sahai_quasimonoenergetic_2018}%
  \BibitemOpen
  \bibfield  {author} {\bibinfo {author} {\bibfnamefont {A.~A.}\ \bibnamefont {Sahai}},\ }\href {https://doi.org/10.1103/PhysRevAccelBeams.21.081301} {\bibfield  {journal} {\bibinfo  {journal} {Physical Review Accelerators and Beams}\ }\textbf {\bibinfo {volume} {21}},\ \bibinfo {pages} {081301} (\bibinfo {year} {2018})}\BibitemShut {NoStop}%
\bibitem [{\citenamefont {Agostinelli}\ \emph {et~al.}(2003)\citenamefont {Agostinelli}, \citenamefont {Allison}, \citenamefont {Amako}, \citenamefont {Apostolakis}, \citenamefont {Araujo}, \citenamefont {Arce}, \citenamefont {Asai}, \citenamefont {Axen}, \citenamefont {Banerjee}, \citenamefont {Barrand},\ and\ \citenamefont {{others}}}]{agostinelli_geant4simulation_2003}%
  \BibitemOpen
  \bibfield  {author} {\bibinfo {author} {\bibfnamefont {S.}~\bibnamefont {Agostinelli}}, \bibinfo {author} {\bibfnamefont {J.}~\bibnamefont {Allison}}, \bibinfo {author} {\bibfnamefont {K.~a.}\ \bibnamefont {Amako}}, \bibinfo {author} {\bibfnamefont {J.}~\bibnamefont {Apostolakis}}, \bibinfo {author} {\bibfnamefont {H.}~\bibnamefont {Araujo}}, \bibinfo {author} {\bibfnamefont {P.}~\bibnamefont {Arce}}, \bibinfo {author} {\bibfnamefont {M.}~\bibnamefont {Asai}}, \bibinfo {author} {\bibfnamefont {D.}~\bibnamefont {Axen}}, \bibinfo {author} {\bibfnamefont {S.}~\bibnamefont {Banerjee}}, \bibinfo {author} {\bibfnamefont {G.}~\bibnamefont {Barrand}},\ and\ \bibinfo {author} {\bibnamefont {{others}}},\ }\href@noop {} {\bibfield  {journal} {\bibinfo  {journal} {Nuclear instruments and methods in physics research section A: Accelerators, Spectrometers, Detectors and Associated Equipment}\ }\textbf {\bibinfo {volume} {506}},\ \bibinfo {pages} {250} (\bibinfo {year} {2003})},\ \bibinfo {note} {publisher: Elsevier}\BibitemShut {NoStop}%
\bibitem [{\citenamefont {Allison}\ \emph {et~al.}(2006)\citenamefont {Allison}, \citenamefont {Amako}, \citenamefont {Apostolakis}, \citenamefont {Araujo}, \citenamefont {Arce~Dubois}, \citenamefont {Asai}, \citenamefont {Barrand}, \citenamefont {Capra}, \citenamefont {Chauvie}, \citenamefont {Chytracek}, \citenamefont {Cirrone}, \citenamefont {Cooperman}, \citenamefont {Cosmo}, \citenamefont {Cuttone}, \citenamefont {Daquino}, \citenamefont {Donszelmann}, \citenamefont {Dressel}, \citenamefont {Folger}, \citenamefont {Foppiano}, \citenamefont {Generowicz}, \citenamefont {Grichine}, \citenamefont {Guatelli}, \citenamefont {Gumplinger}, \citenamefont {Heikkinen}, \citenamefont {Hrivnacova}, \citenamefont {Howard}, \citenamefont {Incerti}, \citenamefont {Ivanchenko}, \citenamefont {Johnson}, \citenamefont {Jones}, \citenamefont {Koi}, \citenamefont {Kokoulin}, \citenamefont {Kossov}, \citenamefont {Kurashige}, \citenamefont {Lara}, \citenamefont {Larsson}, \citenamefont {Lei}, \citenamefont {Link}, \citenamefont {Longo}, \citenamefont {Maire}, \citenamefont {Mantero}, \citenamefont {Mascialino}, \citenamefont {McLaren}, \citenamefont {Mendez~Lorenzo}, \citenamefont {Minamimoto}, \citenamefont {Murakami}, \citenamefont {Nieminen}, \citenamefont {Pandola}, \citenamefont {Parlati}, \citenamefont {Peralta}, \citenamefont {Perl}, \citenamefont {Pfeiffer}, \citenamefont {Pia}, \citenamefont {Ribon}, \citenamefont {Rodrigues}, \citenamefont {Russo}, \citenamefont {Sadilov}, \citenamefont {Santin}, \citenamefont {Sasaki}, \citenamefont {Smith}, \citenamefont {Starkov}, \citenamefont {Tanaka}, \citenamefont {Tcherniaev}, \citenamefont {Tome}, \citenamefont {Trindade}, \citenamefont {Truscott}, \citenamefont {Urban}, \citenamefont {Verderi}, \citenamefont {Walkden}, \citenamefont {Wellisch}, \citenamefont {Williams}, \citenamefont {Wright},\ and\ \citenamefont {Yoshida}}]{allison_geant4_2006}%
  \BibitemOpen
  \bibfield  {author} {\bibinfo {author} {\bibfnamefont {J.}~\bibnamefont {Allison}}, \bibinfo {author} {\bibfnamefont {K.}~\bibnamefont {Amako}}, \bibinfo {author} {\bibfnamefont {J.}~\bibnamefont {Apostolakis}}, \bibinfo {author} {\bibfnamefont {H.}~\bibnamefont {Araujo}}, \bibinfo {author} {\bibfnamefont {P.}~\bibnamefont {Arce~Dubois}}, \bibinfo {author} {\bibfnamefont {M.}~\bibnamefont {Asai}}, \bibinfo {author} {\bibfnamefont {G.}~\bibnamefont {Barrand}}, \bibinfo {author} {\bibfnamefont {R.}~\bibnamefont {Capra}}, \bibinfo {author} {\bibfnamefont {S.}~\bibnamefont {Chauvie}}, \bibinfo {author} {\bibfnamefont {R.}~\bibnamefont {Chytracek}}, \bibinfo {author} {\bibfnamefont {G.}~\bibnamefont {Cirrone}}, \bibinfo {author} {\bibfnamefont {G.}~\bibnamefont {Cooperman}}, \bibinfo {author} {\bibfnamefont {G.}~\bibnamefont {Cosmo}}, \bibinfo {author} {\bibfnamefont {G.}~\bibnamefont {Cuttone}}, \bibinfo {author} {\bibfnamefont {G.}~\bibnamefont {Daquino}}, \bibinfo {author} {\bibfnamefont {M.}~\bibnamefont {Donszelmann}}, \bibinfo {author} {\bibfnamefont {M.}~\bibnamefont {Dressel}}, \bibinfo {author} {\bibfnamefont {G.}~\bibnamefont {Folger}}, \bibinfo {author} {\bibfnamefont {F.}~\bibnamefont {Foppiano}}, \bibinfo {author} {\bibfnamefont {J.}~\bibnamefont {Generowicz}}, \bibinfo {author} {\bibfnamefont {V.}~\bibnamefont {Grichine}}, \bibinfo {author} {\bibfnamefont {S.}~\bibnamefont {Guatelli}}, \bibinfo {author} {\bibfnamefont {P.}~\bibnamefont {Gumplinger}}, \bibinfo {author} {\bibfnamefont {A.}~\bibnamefont {Heikkinen}}, \bibinfo {author} {\bibfnamefont {I.}~\bibnamefont {Hrivnacova}}, \bibinfo {author} {\bibfnamefont {A.}~\bibnamefont {Howard}}, \bibinfo {author} {\bibfnamefont {S.}~\bibnamefont {Incerti}}, \bibinfo {author} {\bibfnamefont {V.}~\bibnamefont {Ivanchenko}}, \bibinfo {author} {\bibfnamefont {T.}~\bibnamefont {Johnson}}, \bibinfo {author} {\bibfnamefont {F.}~\bibnamefont {Jones}}, \bibinfo {author} {\bibfnamefont {T.}~\bibnamefont {Koi}}, \bibinfo {author} {\bibfnamefont {R.}~\bibnamefont {Kokoulin}}, \bibinfo {author} {\bibfnamefont {M.}~\bibnamefont {Kossov}}, \bibinfo {author} {\bibfnamefont {H.}~\bibnamefont {Kurashige}}, \bibinfo {author} {\bibfnamefont {V.}~\bibnamefont {Lara}}, \bibinfo {author} {\bibfnamefont {S.}~\bibnamefont {Larsson}}, \bibinfo {author} {\bibfnamefont {F.}~\bibnamefont {Lei}}, \bibinfo {author} {\bibfnamefont {O.}~\bibnamefont {Link}}, \bibinfo {author} {\bibfnamefont {F.}~\bibnamefont {Longo}}, \bibinfo {author} {\bibfnamefont {M.}~\bibnamefont {Maire}}, \bibinfo {author} {\bibfnamefont {A.}~\bibnamefont {Mantero}}, \bibinfo {author} {\bibfnamefont {B.}~\bibnamefont {Mascialino}}, \bibinfo {author} {\bibfnamefont {I.}~\bibnamefont {McLaren}}, \bibinfo {author} {\bibfnamefont {P.}~\bibnamefont {Mendez~Lorenzo}}, \bibinfo {author} {\bibfnamefont {K.}~\bibnamefont {Minamimoto}}, \bibinfo {author} {\bibfnamefont {K.}~\bibnamefont {Murakami}}, \bibinfo {author} {\bibfnamefont {P.}~\bibnamefont {Nieminen}}, \bibinfo {author} {\bibfnamefont {L.}~\bibnamefont {Pandola}}, \bibinfo {author} {\bibfnamefont {S.}~\bibnamefont {Parlati}}, \bibinfo {author} {\bibfnamefont {L.}~\bibnamefont {Peralta}}, \bibinfo {author} {\bibfnamefont {J.}~\bibnamefont {Perl}}, \bibinfo {author} {\bibfnamefont {A.}~\bibnamefont {Pfeiffer}}, \bibinfo {author} {\bibfnamefont {M.}~\bibnamefont {Pia}}, \bibinfo {author} {\bibfnamefont {A.}~\bibnamefont {Ribon}}, \bibinfo {author} {\bibfnamefont {P.}~\bibnamefont {Rodrigues}}, \bibinfo {author} {\bibfnamefont {G.}~\bibnamefont {Russo}}, \bibinfo {author} {\bibfnamefont {S.}~\bibnamefont {Sadilov}}, \bibinfo {author} {\bibfnamefont {G.}~\bibnamefont {Santin}}, \bibinfo {author} {\bibfnamefont {T.}~\bibnamefont {Sasaki}}, \bibinfo {author} {\bibfnamefont {D.}~\bibnamefont {Smith}}, \bibinfo {author} {\bibfnamefont {N.}~\bibnamefont {Starkov}}, \bibinfo {author} {\bibfnamefont {S.}~\bibnamefont {Tanaka}}, \bibinfo {author} {\bibfnamefont {E.}~\bibnamefont {Tcherniaev}}, \bibinfo {author} {\bibfnamefont {B.}~\bibnamefont {Tome}}, \bibinfo {author} {\bibfnamefont {A.}~\bibnamefont {Trindade}}, \bibinfo {author} {\bibfnamefont {P.}~\bibnamefont {Truscott}}, \bibinfo {author} {\bibfnamefont {L.}~\bibnamefont {Urban}}, \bibinfo {author} {\bibfnamefont {M.}~\bibnamefont {Verderi}}, \bibinfo {author} {\bibfnamefont {A.}~\bibnamefont {Walkden}}, \bibinfo {author} {\bibfnamefont {J.}~\bibnamefont {Wellisch}}, \bibinfo {author} {\bibfnamefont {D.}~\bibnamefont {Williams}}, \bibinfo {author} {\bibfnamefont {D.}~\bibnamefont {Wright}},\ and\ \bibinfo {author} {\bibfnamefont {H.}~\bibnamefont {Yoshida}},\ }\href {https://doi.org/10.1109/TNS.2006.869826} {\bibfield  {journal} {\bibinfo  {journal} {IEEE Transactions on Nuclear Science}\ }\textbf {\bibinfo {volume} {53}},\ \bibinfo {pages} {270} (\bibinfo {year} {2006})}\BibitemShut {NoStop}%
\bibitem [{\citenamefont {Allison}\ \emph {et~al.}(2016)\citenamefont {Allison}, \citenamefont {Amako}, \citenamefont {Apostolakis}, \citenamefont {Arce}, \citenamefont {Asai}, \citenamefont {Aso}, \citenamefont {Bagli}, \citenamefont {Bagulya}, \citenamefont {Banerjee}, \citenamefont {Barrand},\ and\ \citenamefont {{others}}}]{allison_recent_2016}%
  \BibitemOpen
  \bibfield  {author} {\bibinfo {author} {\bibfnamefont {J.}~\bibnamefont {Allison}}, \bibinfo {author} {\bibfnamefont {K.}~\bibnamefont {Amako}}, \bibinfo {author} {\bibfnamefont {J.}~\bibnamefont {Apostolakis}}, \bibinfo {author} {\bibfnamefont {P.}~\bibnamefont {Arce}}, \bibinfo {author} {\bibfnamefont {M.}~\bibnamefont {Asai}}, \bibinfo {author} {\bibfnamefont {T.}~\bibnamefont {Aso}}, \bibinfo {author} {\bibfnamefont {E.}~\bibnamefont {Bagli}}, \bibinfo {author} {\bibfnamefont {A.}~\bibnamefont {Bagulya}}, \bibinfo {author} {\bibfnamefont {S.}~\bibnamefont {Banerjee}}, \bibinfo {author} {\bibfnamefont {G.}~\bibnamefont {Barrand}},\ and\ \bibinfo {author} {\bibnamefont {{others}}},\ }\href@noop {} {\bibfield  {journal} {\bibinfo  {journal} {Nuclear instruments and methods in physics research section A: Accelerators, Spectrometers, Detectors and Associated Equipment}\ }\textbf {\bibinfo {volume} {835}},\ \bibinfo {pages} {186} (\bibinfo {year} {2016})},\ \bibinfo {note} {publisher: Elsevier}\BibitemShut {NoStop}%
\bibitem [{\citenamefont {Benedetti}\ \emph {et~al.}(2018)\citenamefont {Benedetti}, \citenamefont {Schroeder}, \citenamefont {Mehrling}, \citenamefont {Djordjevic}, \citenamefont {Bulanov}, \citenamefont {Geddes}, \citenamefont {Esarey},\ and\ \citenamefont {Leemans}}]{benedetti_infrno_2018}%
  \BibitemOpen
  \bibfield  {author} {\bibinfo {author} {\bibfnamefont {C.}~\bibnamefont {Benedetti}}, \bibinfo {author} {\bibfnamefont {C.}~\bibnamefont {Schroeder}}, \bibinfo {author} {\bibfnamefont {T.}~\bibnamefont {Mehrling}}, \bibinfo {author} {\bibfnamefont {B.}~\bibnamefont {Djordjevic}}, \bibinfo {author} {\bibfnamefont {S.}~\bibnamefont {Bulanov}}, \bibinfo {author} {\bibfnamefont {C.}~\bibnamefont {Geddes}}, \bibinfo {author} {\bibfnamefont {E.}~\bibnamefont {Esarey}},\ and\ \bibinfo {author} {\bibfnamefont {W.}~\bibnamefont {Leemans}},\ }in\ \href {https://doi.org/10.1109/AAC.2018.8659411} {\emph {\bibinfo {booktitle} {2018 {IEEE} {Advanced} {Accelerator} {Concepts} {Workshop} ({AAC})}}}\ (\bibinfo {year} {2018})\ pp.\ \bibinfo {pages} {1--5}\BibitemShut {NoStop}%
\bibitem [{\citenamefont {Gryaznykh}\ \emph {et~al.}(1998)\citenamefont {Gryaznykh}, \citenamefont {Kandiev},\ and\ \citenamefont {Lykov}}]{gryaznykh_estimates_1998}%
  \BibitemOpen
  \bibfield  {author} {\bibinfo {author} {\bibfnamefont {D.~A.}\ \bibnamefont {Gryaznykh}}, \bibinfo {author} {\bibfnamefont {Y.~Z.}\ \bibnamefont {Kandiev}},\ and\ \bibinfo {author} {\bibfnamefont {V.~A.}\ \bibnamefont {Lykov}},\ }\href {https://doi.org/10.1134/1.567660} {\bibfield  {journal} {\bibinfo  {journal} {Journal of Experimental and Theoretical Physics Letters}\ }\textbf {\bibinfo {volume} {67}},\ \bibinfo {pages} {257} (\bibinfo {year} {1998})}\BibitemShut {NoStop}%
\bibitem [{\citenamefont {Nakashima}\ and\ \citenamefont {Takabe}(2002)}]{nakashima_numerical_2002}%
  \BibitemOpen
  \bibfield  {author} {\bibinfo {author} {\bibfnamefont {K.}~\bibnamefont {Nakashima}}\ and\ \bibinfo {author} {\bibfnamefont {H.}~\bibnamefont {Takabe}},\ }\href {https://doi.org/10.1063/1.1464145} {\bibfield  {journal} {\bibinfo  {journal} {Physics of Plasmas}\ }\textbf {\bibinfo {volume} {9}},\ \bibinfo {pages} {1505} (\bibinfo {year} {2002})}\BibitemShut {NoStop}%
\bibitem [{\citenamefont {Scott}\ \emph {et~al.}(2015)\citenamefont {Scott}, \citenamefont {Bagnoud}, \citenamefont {Brabetz}, \citenamefont {Clarke}, \citenamefont {Green}, \citenamefont {Heathcote}, \citenamefont {Powell}, \citenamefont {Zielbauer}, \citenamefont {Arber}, \citenamefont {McKenna},\ and\ \citenamefont {{others}}}]{scott_optimization_2015}%
  \BibitemOpen
  \bibfield  {author} {\bibinfo {author} {\bibfnamefont {G.}~\bibnamefont {Scott}}, \bibinfo {author} {\bibfnamefont {V.}~\bibnamefont {Bagnoud}}, \bibinfo {author} {\bibfnamefont {C.}~\bibnamefont {Brabetz}}, \bibinfo {author} {\bibfnamefont {R.}~\bibnamefont {Clarke}}, \bibinfo {author} {\bibfnamefont {J.}~\bibnamefont {Green}}, \bibinfo {author} {\bibfnamefont {R.}~\bibnamefont {Heathcote}}, \bibinfo {author} {\bibfnamefont {H.}~\bibnamefont {Powell}}, \bibinfo {author} {\bibfnamefont {B.}~\bibnamefont {Zielbauer}}, \bibinfo {author} {\bibfnamefont {T.}~\bibnamefont {Arber}}, \bibinfo {author} {\bibfnamefont {P.}~\bibnamefont {McKenna}},\ and\ \bibinfo {author} {\bibnamefont {{others}}},\ }\href@noop {} {\bibfield  {journal} {\bibinfo  {journal} {New Journal of Physics}\ }\textbf {\bibinfo {volume} {17}},\ \bibinfo {pages} {033027} (\bibinfo {year} {2015})},\ \bibinfo {note} {publisher: IOP Publishing}\BibitemShut {NoStop}%
\bibitem [{\citenamefont {Gorbunov}\ and\ \citenamefont {Kirsanov}(1987)}]{gorbunov_excitation_1987}%
  \BibitemOpen
  \bibfield  {author} {\bibinfo {author} {\bibfnamefont {L.~M.}\ \bibnamefont {Gorbunov}}\ and\ \bibinfo {author} {\bibfnamefont {V.~I.}\ \bibnamefont {Kirsanov}},\ }\href@noop {} {\bibfield  {journal} {\bibinfo  {journal} {Zh. Eksp. Teor. Fiz}\ } (\bibinfo {year} {1987})}\BibitemShut {NoStop}%
\bibitem [{\citenamefont {Diederichs}\ \emph {et~al.}(2019)\citenamefont {Diederichs}, \citenamefont {Mehrling}, \citenamefont {Benedetti}, \citenamefont {Schroeder}, \citenamefont {Knetsch}, \citenamefont {Esarey},\ and\ \citenamefont {Osterhoff}}]{diederichs_positron_2019}%
  \BibitemOpen
  \bibfield  {author} {\bibinfo {author} {\bibfnamefont {S.}~\bibnamefont {Diederichs}}, \bibinfo {author} {\bibfnamefont {T.~J.}\ \bibnamefont {Mehrling}}, \bibinfo {author} {\bibfnamefont {C.}~\bibnamefont {Benedetti}}, \bibinfo {author} {\bibfnamefont {C.~B.}\ \bibnamefont {Schroeder}}, \bibinfo {author} {\bibfnamefont {A.}~\bibnamefont {Knetsch}}, \bibinfo {author} {\bibfnamefont {E.}~\bibnamefont {Esarey}},\ and\ \bibinfo {author} {\bibfnamefont {J.}~\bibnamefont {Osterhoff}},\ }\href {https://doi.org/10.1103/PhysRevAccelBeams.22.081301} {\bibfield  {journal} {\bibinfo  {journal} {Physical Review Accelerators and Beams}\ }\textbf {\bibinfo {volume} {22}},\ \bibinfo {pages} {081301} (\bibinfo {year} {2019})},\ \bibinfo {note} {publisher: American Physical Society}\BibitemShut {NoStop}%
\bibitem [{\citenamefont {Benedetti}\ \emph {et~al.}(2010)\citenamefont {Benedetti}, \citenamefont {Schroeder}, \citenamefont {Esarey}, \citenamefont {Geddes},\ and\ \citenamefont {Leemans}}]{benedetti_efficient_2010}%
  \BibitemOpen
  \bibfield  {author} {\bibinfo {author} {\bibfnamefont {C.}~\bibnamefont {Benedetti}}, \bibinfo {author} {\bibfnamefont {C.~B.}\ \bibnamefont {Schroeder}}, \bibinfo {author} {\bibfnamefont {E.}~\bibnamefont {Esarey}}, \bibinfo {author} {\bibfnamefont {C.~G.~R.}\ \bibnamefont {Geddes}},\ and\ \bibinfo {author} {\bibfnamefont {W.~P.}\ \bibnamefont {Leemans}},\ }\href {https://doi.org/10.1063/1.3520323} {\bibfield  {journal} {\bibinfo  {journal} {AIP Conference Proceedings}\ }\textbf {\bibinfo {volume} {1299}},\ \bibinfo {pages} {250} (\bibinfo {year} {2010})},\ \bibinfo {note} {publisher: American Institute of Physics}\BibitemShut {NoStop}%
\bibitem [{\citenamefont {Benedetti}\ \emph {et~al.}(2017)\citenamefont {Benedetti}, \citenamefont {Schroeder}, \citenamefont {Geddes}, \citenamefont {Esarey},\ and\ \citenamefont {Leemans}}]{benedetti_accurate_2017}%
  \BibitemOpen
  \bibfield  {author} {\bibinfo {author} {\bibfnamefont {C.}~\bibnamefont {Benedetti}}, \bibinfo {author} {\bibfnamefont {C.~B.}\ \bibnamefont {Schroeder}}, \bibinfo {author} {\bibfnamefont {C.~G.~R.}\ \bibnamefont {Geddes}}, \bibinfo {author} {\bibfnamefont {E.}~\bibnamefont {Esarey}},\ and\ \bibinfo {author} {\bibfnamefont {W.~P.}\ \bibnamefont {Leemans}},\ }\href {https://doi.org/10.1088/1361-6587/aa8977} {\bibfield  {journal} {\bibinfo  {journal} {Plasma Physics and Controlled Fusion}\ }\textbf {\bibinfo {volume} {60}},\ \bibinfo {pages} {014002} (\bibinfo {year} {2017})},\ \bibinfo {note} {publisher: IOP Publishing}\BibitemShut {NoStop}%
\bibitem [{\citenamefont {Benedetti}\ \emph {et~al.}(2015)\citenamefont {Benedetti}, \citenamefont {Rossi}, \citenamefont {Schroeder}, \citenamefont {Esarey},\ and\ \citenamefont {Leemans}}]{benedetti_pulse_2015}%
  \BibitemOpen
  \bibfield  {author} {\bibinfo {author} {\bibfnamefont {C.}~\bibnamefont {Benedetti}}, \bibinfo {author} {\bibfnamefont {F.}~\bibnamefont {Rossi}}, \bibinfo {author} {\bibfnamefont {C.~B.}\ \bibnamefont {Schroeder}}, \bibinfo {author} {\bibfnamefont {E.}~\bibnamefont {Esarey}},\ and\ \bibinfo {author} {\bibfnamefont {W.~P.}\ \bibnamefont {Leemans}},\ }\href {https://doi.org/10.1103/PhysRevE.92.023109} {\bibfield  {journal} {\bibinfo  {journal} {Physical Review E}\ }\textbf {\bibinfo {volume} {92}},\ \bibinfo {pages} {023109} (\bibinfo {year} {2015})}\BibitemShut {NoStop}%
\end{thebibliography}%

\end{document}